\documentclass [11pt, letterpaper]{article}
\usepackage{fullpage}
\usepackage{amsmath}
\usepackage{amssymb}
\usepackage{graphicx}
\usepackage{mathrsfs}
\usepackage{wrapfig}
\usepackage{boxedminipage}
\usepackage{epsfig}
\usepackage{setspace}
\usepackage{subfigure}
\usepackage{float}
\usepackage{hyperref}
\usepackage{enumerate}

\begin{document}

\begin{flushright}
{\tt arXiv:1306.4955}
\end{flushright}

{\flushleft\vskip-1.35cm\vbox{\includegraphics[width=1.25in]{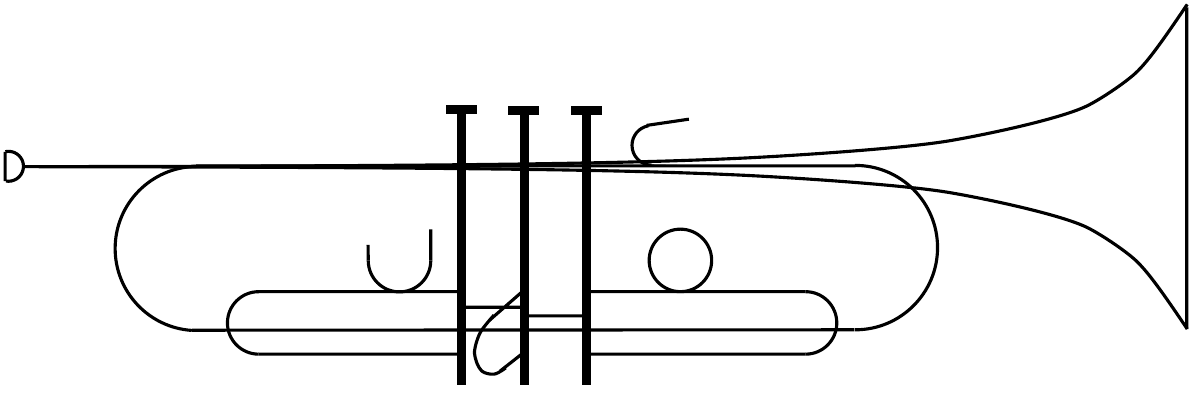}}}

\bigskip
\bigskip
\bigskip

\begin{center} 

{\Large\bf  Large $N$ Phase Transitions, }

\bigskip
\bigskip

{\Large\bf   Finite Volume,}

\bigskip
\bigskip

{\Large\bf   and}

\bigskip
\bigskip

{\Large\bf    Entanglement Entropy }

\end{center}

\bigskip \bigskip \bigskip \bigskip

\centerline{\bf Clifford V. Johnson}

\bigskip
\bigskip
\bigskip

  \centerline{\it Department of Physics and Astronomy }
\centerline{\it University of
Southern California}
\centerline{\it Los Angeles, CA 90089-0484, U.S.A.}

\bigskip

\centerline{\small \tt johnson1,  [at] usc.edu}

\bigskip
\bigskip


\begin{abstract} 
\noindent 
Holographic studies of the entanglement entropy of field theories dual to charged and neutral black holes in asymptotically global AdS$_4$ spacetimes are presented. The goal is to elucidate various properties of the quantity that are peculiar to working in finite volume, and to gain access to the behaviour of the entanglement entropy in the  rich thermodynamic phase structure that is present at finite volume and large~$N$. The entropy is followed through various first order phase transitions, and also a novel second order phase transition. Properties of a specific heat in the neighbourhood of the critical point are computed. Behaviour is found that contrasts interestingly with an earlier holographic study of  a second order phase transition dual to an holographic superconductor.

\end{abstract}
\newpage \baselineskip=18pt \setcounter{footnote}{0}

\section{Introduction}
%

There are a number of reasons to consider the nature of entanglement entropy in systems of finite volume. Most obviously, perhaps, is the fact that since the entanglement entropy's definition is so geometrical in nature (see section~\ref{sec:entanglement} for a reminder), important new features are sure to arise as a result of the finiteness of {\it both} regions whose mutual entanglement are being considered, as well as new considerations about the nature of how the entangling regions' relative shape are chosen. Some of these features will emerge  in the results presented, but the main motivation  was to use finite volume as a laboratory for studying the entanglement entropy in situations where there is interesting phase structure. Genuine thermodynamic phase structure is possible in finite volume if there is some other way of getting the needed thermodynamic limit that comes from having an infinite number of degrees of freedom. 

One such way is to be able to dial up the number of degrees of freedom by having multi--component fields such as vectors, matrices, and so forth, allowing one to send the degrees of freedom to infinity in an organized fashion. Gauge theories (and close cousins thereof)  are a good example, and the large $N$ limit (where there are order $N^2$ fields in the adjoint representation of the gauge group, for example) has long\cite{'tHooft:1973jz} been known as a famous way of yielding interesting models. Holographic studies of gauge theories at large $N$, in the form of the AdS/CFT correspondence\cite{Maldacena:1997re, Gubser:1998bc, Witten:1998qj, Witten:1998zw} and generalizations, have yielded a large number of interesting examples of phase transitions in both finite and infinite volume, and these examples have been probed in a number of ways, providing considerable insights in a number of directions. With on one hand, the (conjectured) holographic definition of the entanglement entropy  of Ryu and Takayanagi\cite{Ryu:2006bv,Ryu:2006ef}, and  on the other hand the growing interest in understanding entanglement entropy in a wide range of applications  to quantum systems, it seems prudent to see what can be learned by revisiting some old familiar and robust examples, tracking the entanglement entropy's properties where possible. 

That is the goal of the work being reported here, and the target examples (focusing on finite volume) will be asymptotically anti--de Sitter charged and uncharged black holes  in Einstein--Maxwell gravity, for which much of the phase structure and its holographic interpretations were uncovered in refs.\cite{Hawking:1982dh,Witten:1998zw,Chamblin:1999tk}, as we will review below. These examples afford us first order finite temperature transitions, an understanding of how to characterize the phases on either side of the transition (in terms of properties of the black hole, or of AdS), novel first order transitions between different types of black hole, and (even more novel perhaps) a second order phase transition at the end of a critical line of first order points. This is all rich physics which deserves to be revisited given the new ability to get access to the entanglement entropy and track it across the transition. 

This is all very interesting in its own right, but there's a second agendum here. We need to enlarge the number of tractable examples in which we have computational control of the entanglement  entropy, especially if it is to be used as a diagnostic tool in more complicated systems of both theoretical and experimental interest, such as novel condensed matter phases arising as a result of thermal or quantum phase transitions. For example, the second order point accessible at finite volume could well teach us some interesting physics that may have broader lessons, allowing us to compare and contrast with the physics of other second order points, or to deduce physics about points to which we do not have ready computational access.  Broadly speaking, this paper's results may be considered  an attempt to widen the class of examples we have at our fingertips.

\section{Review of Holographic Entanglement Entropy}
\label{sec:entanglement}

Given a quantum system, the entanglement entropy  of a subsystem $\cal A$ and its complement $\cal B$ has the following  definition\footnote{Ref.\cite{Ryu:2006ef} has a review and several useful references.}:
\begin{equation}
S_{\mathcal{A}} = - \mathrm{Tr}_{\mathcal{A}} \left( \rho_\mathcal{A} \ln \rho_\mathcal{A} \right)\ ,
\end{equation}
where $\rho_\mathcal{A}$ is the reduced density matrix of $\mathcal{A}$ given by tracing over the degrees of freedom of $\mathcal{B}$,
$\rho_\mathcal{A} = \mathrm{Tr}_{\mathcal{B}}( \rho) $,
with $\rho$ is the density matrix of the system.  

When there is a holographically dual gravitational system available for the quantum system that is asymptotically anti--de Sitter (AdS), it has been conjectured that the entropy is holographically computed as follows\cite{Ryu:2006bv,Ryu:2006ef}. Consider a slice at a constant value of the AdS radial
coordinate. Recall that this defines the dual field theory
(with one dimension fewer) as essentially residing on that slice in
the presence of a UV cutoff set by the position of the slice. Sending
the slice to the AdS boundary at infinity removes the cutoff (see
ref.\cite{Aharony:1999t} for a review).  On the slice, consider
a region $\mathcal{A}$. Now find the minimal--area surface
$\gamma_{\mathcal{A}}$ bounded by the perimeter of $\mathcal{A}$ and
that extends into the bulk of the geometry. 
Then the entanglement
entropy of region $\mathcal{A}$ with $\mathcal{B}$ is given by:
\begin{equation}
\label{eq:holographic_entanglement}
S_{\mathcal{A}} = \frac{ \mathrm{Area}(\gamma_{\mathcal{A}})}{4 G} \ ,
\end{equation}
where $G$ is Newton's constant in the dual gravity
theory. 

Note that there is an important refinement of the prescription for the entropy when there is non--trivial topology in the bulk due to the horizon of a black hole. It is discussed at the beginning of section~\ref{sec:eeglobal}. In this paper, we will not be studying regimes where this prescription is needed.

\section{Some Properties of Charged Black Holes in AdS$_4$}
Let us recall some aspects of the system to be studied. The black hole is a Reissner--Nordstr\"om--like solution of the Einstein--Maxwell system with bulk action
\begin{equation}
I=-\frac{1}{16\pi G}\int \! d^4x \sqrt{-g} \left(R-2\Lambda -F^2\right)\ ,
\end{equation}
where $\Lambda=-3/l^2$, the cosmological constant sets a length scale $l$. The black hole has mass and charge set by parameters $m$ and $q$, with metric
\begin{equation}
ds^2 = -V( r)dt^2
+ {dr^2\over V(r)} + r^2 (d\theta^2+\sin^2\theta d\varphi^2)\ ,
\end{equation}
where
\begin{equation}
V( r) = 1-\frac{m}{r}+\frac{q^2}{r^2}+\frac{r^2}{l^2}\ ,
\end{equation}
and  there is a gauge potential 
\begin{equation}
A_t = -\frac{q}{r}+\Phi\ ,
\end{equation}
with $\Phi=q/r_+$,  a constant chosen to have $A$ vanish on the horizon at $r=r_+$, the largest positive real root of $V( r)$.

\begin{figure}[h]
\begin{center}
\subfigure[$\Phi<\Phi_c=1$.]{\includegraphics[width=3.0in]{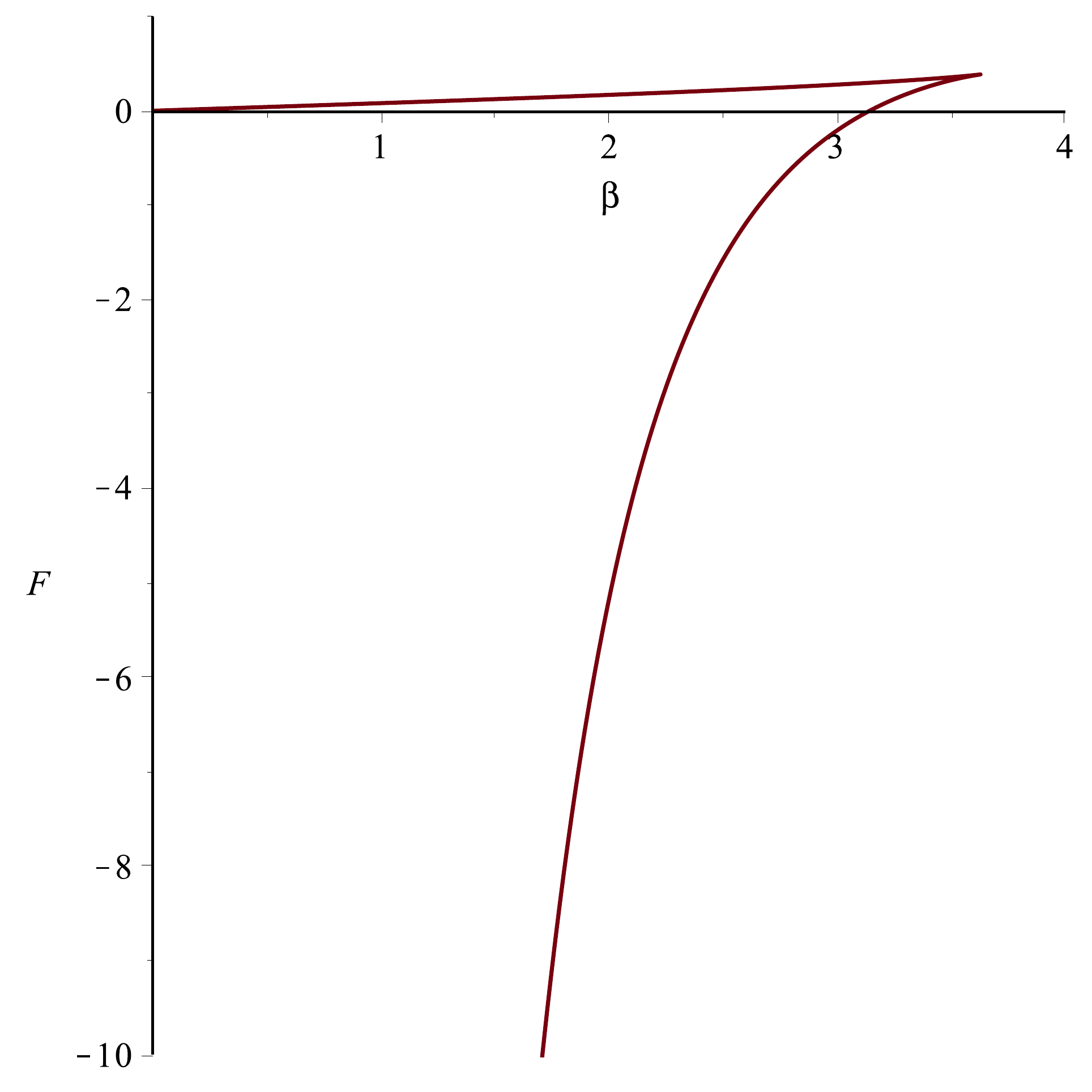}} 
\subfigure[$\Phi\geq\Phi_c=1$.]{\includegraphics[width=3.0in]{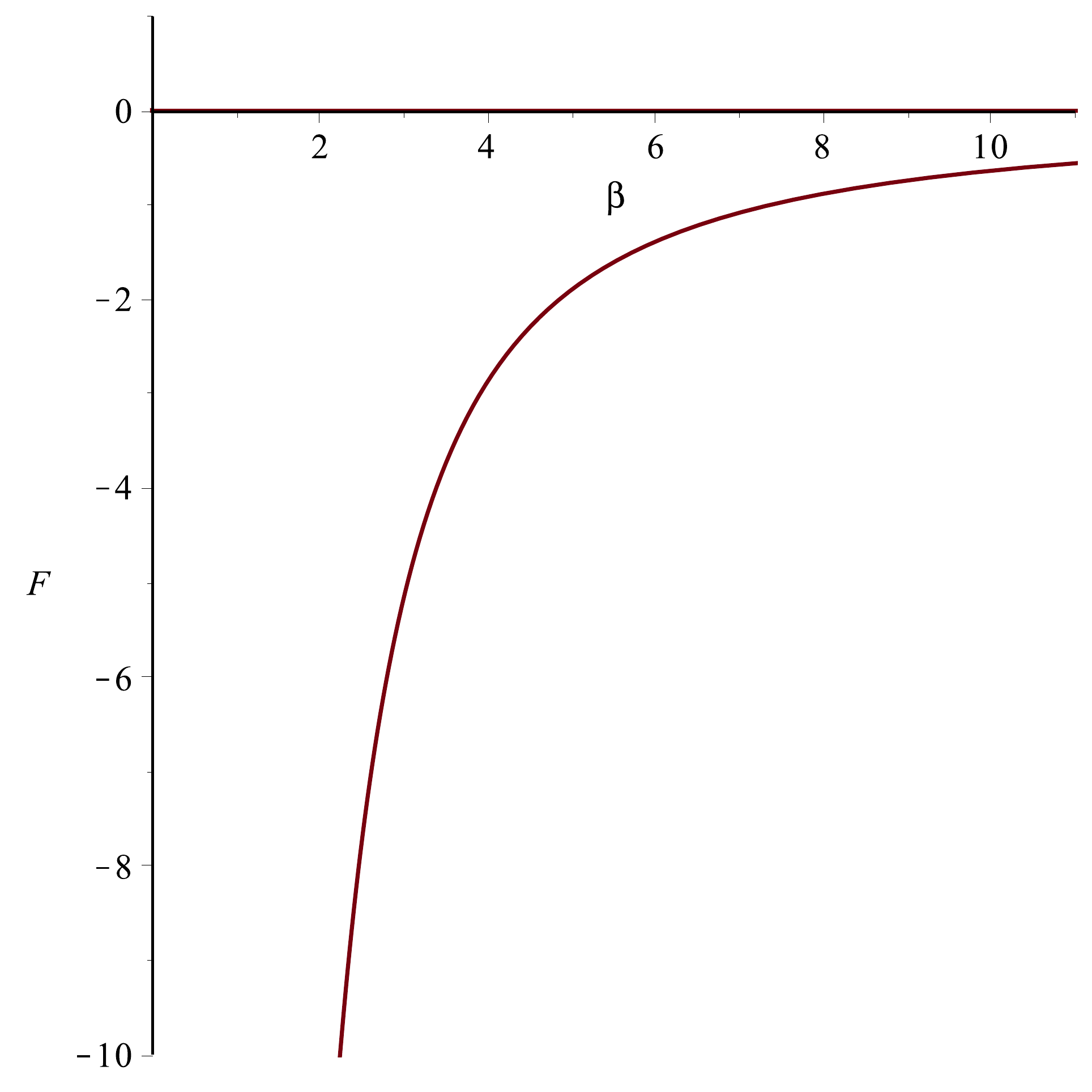}}
   \caption{\small The free energy  {\it vs.} $\beta$ curve in the fixed potential ensemble. For $\Phi<\Phi_c=1$, it is a cusp. Here, and in subsequent figures, we chose $G=1$ and $l=1$ for convenience.}  \label{fig:free_energy1}
   \end{center}
\end{figure}

\begin{figure}[h]
\begin{center}
\subfigure[Shape for $\Phi<\Phi_c=1$.]{\includegraphics[width=3.0in]{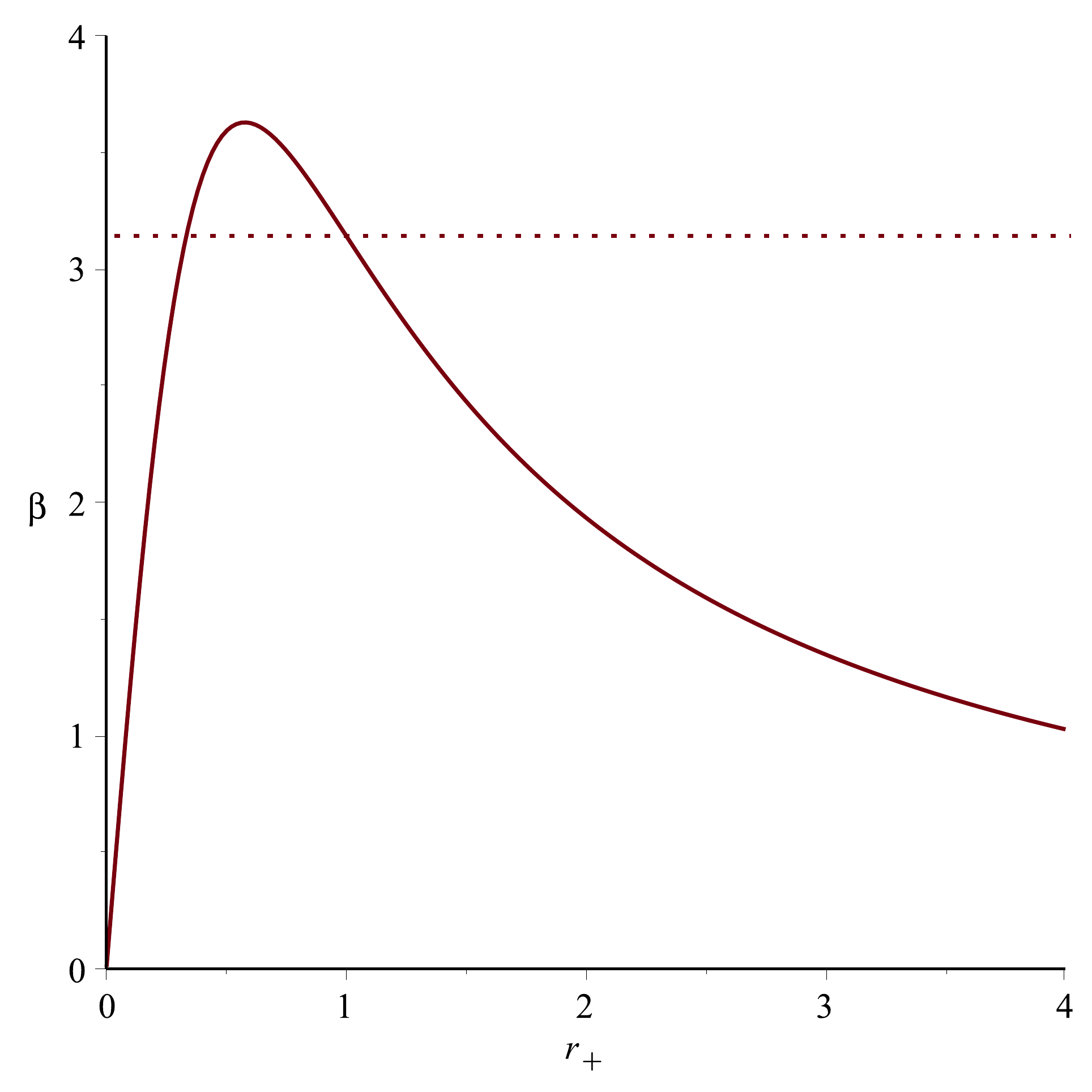}} 
\subfigure[Shape for $\Phi>\Phi_c=1$ or alternatively, $q>q_c=\frac{l}{6}$.]{\includegraphics[width=3.0in]{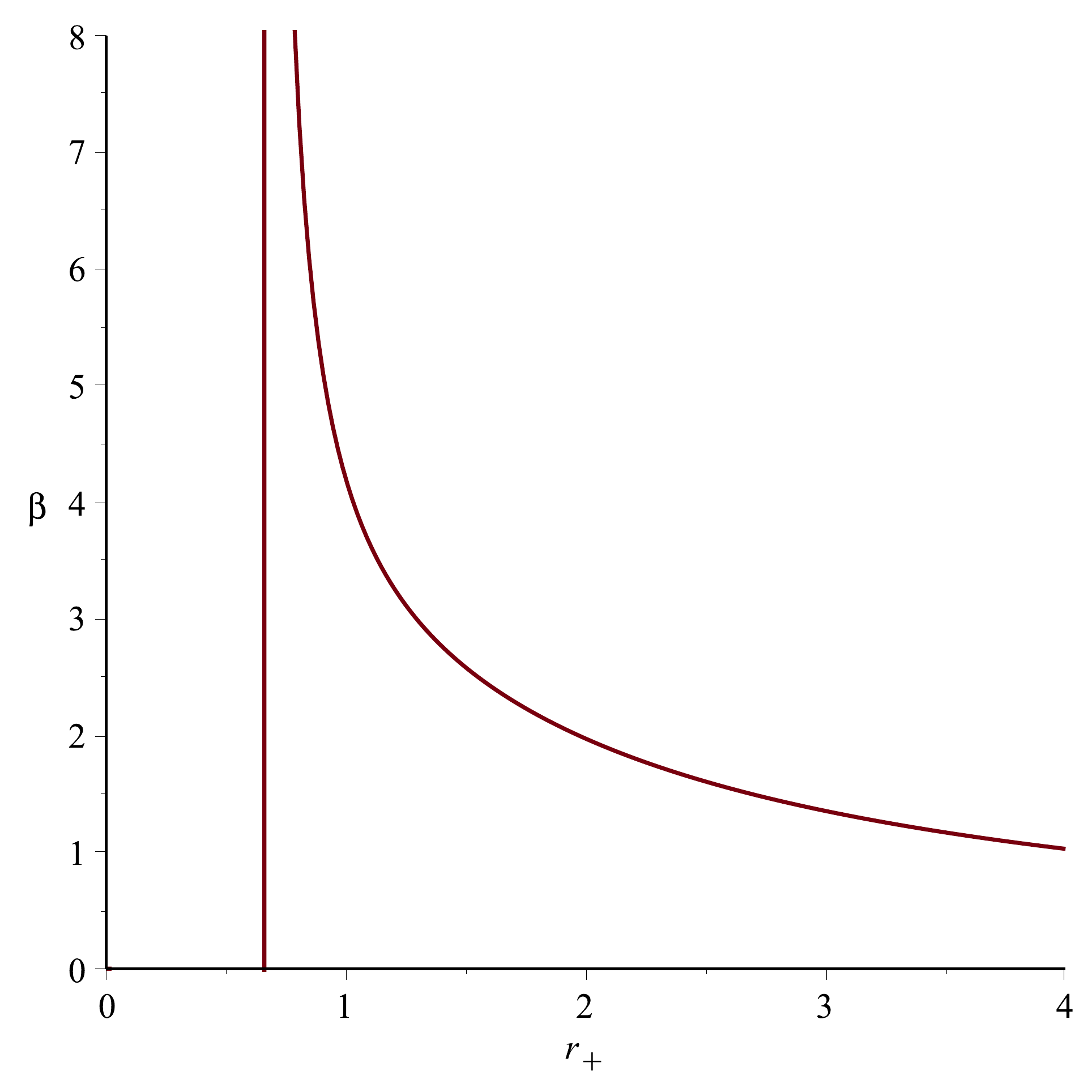}}
   \caption{\small Sample curves showing the shape of the inverse temperature as function of horizon radius for the fixed potential ensemble. The dotted line shows the location of the first order transition. For $\Phi>\Phi_c$, the curve is of roughly the same form as that of the fixed charge ensemble for $q>q_c$. See text for discussion.}  \label{fig:beta_curves1}
   \end{center}
\end{figure}

Various thermodynamic properties of this system were worked out some time ago, starting (for our purposes) with the work of Hawking and Page\cite{Hawking:1982dh},  and greatly clarified and expanded in the context of holographic duality to field theory physics by Witten\cite{Witten:1998zw}.  The extension to Reissner--Nordstr\"om black holes in Einstein--Maxwell--AdS was done in ref.\cite{Chamblin:1999tk}, where several new phenomena were observed, as will be reviewed below. Overall, a key point is that phase transitions can take place as one varies (for example) temperature and charge (or potential, depending upon one's choice of ensemble). From the dual field theory point of view, these are possible even though the system is in finite volume (the dual $(2+1)$--dimensional theory is on $R\times S^2$) because there are still an infinite number of degrees of freedom as a result of being in a large $N$ limit.  At fixed potential $\Phi$, for example, the free energy of the black hole relative to pure AdS is\cite{Chamblin:1999tk}:
\begin{equation}
F=\frac{I}{\beta}=\frac{1}{4 Gl^2}\left( l^2r_+(1-\Phi^2)-r^3_+\right)\  ,
\end{equation}
where the inverse temperature $\beta$ is given by the usual relation
\begin{equation}
\beta=4\pi V^\prime \left.\right|_{r=r_+} = \frac{4\pi l^2 r_+^3}{3r_+^4+l^2r_+^2-q^2 l^2} =  \frac{4\pi l^2 r_+}{3r_+^2+l^2(1-\Phi^2)} \ .
\label{eq:equationofstate}
\end{equation} This is plotted as a function of inverse temperature in figure~\ref{fig:free_energy1}.
The first order transition of the Hawking--Page  system is the special case of $\Phi=0$, and we have a generalization to a family of first order transitions in the $(T,\Phi)$ plane, as worked out in ref.\cite{Chamblin:1999tk}. There are up to three competing phases, depending upon one's location in the plane. For $\Phi<\Phi_c=1$, at large enough temperature,  there is pure AdS, and two families of black hole, ``small" and ``large". The black hole families can be seen by plotting the $\beta$ ``equation of state'' for fixed $\Phi<1$. See figure~\ref{fig:beta_curves1}(a). 
\begin{figure}[h]
\begin{center}
\subfigure[Shape for $q<q_c=\frac{l}{6}$.]{\includegraphics[width=3.0in]{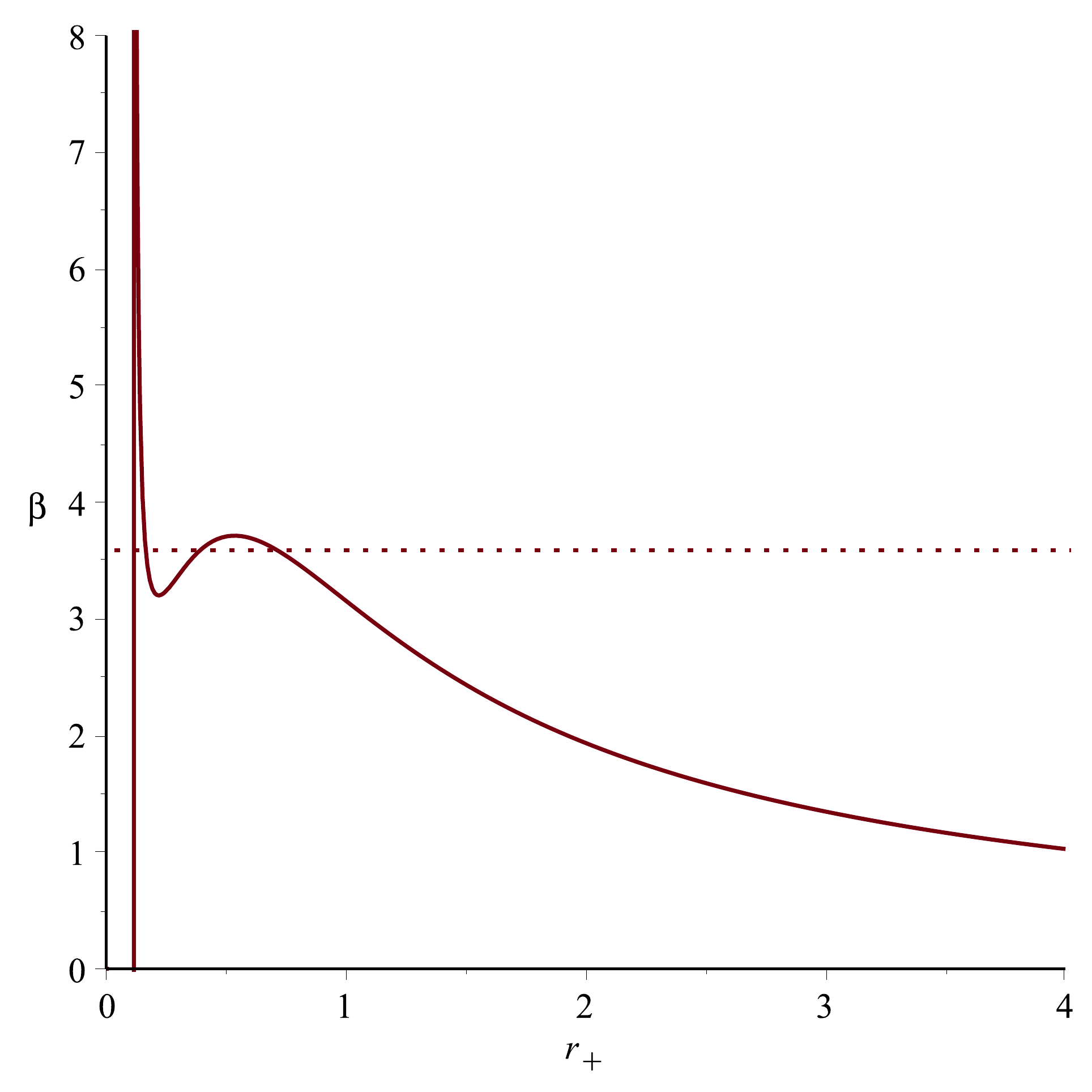}} 
\subfigure[Shape for $q=q_c=\frac{l}{6}$.]{\includegraphics[width=3.0in]{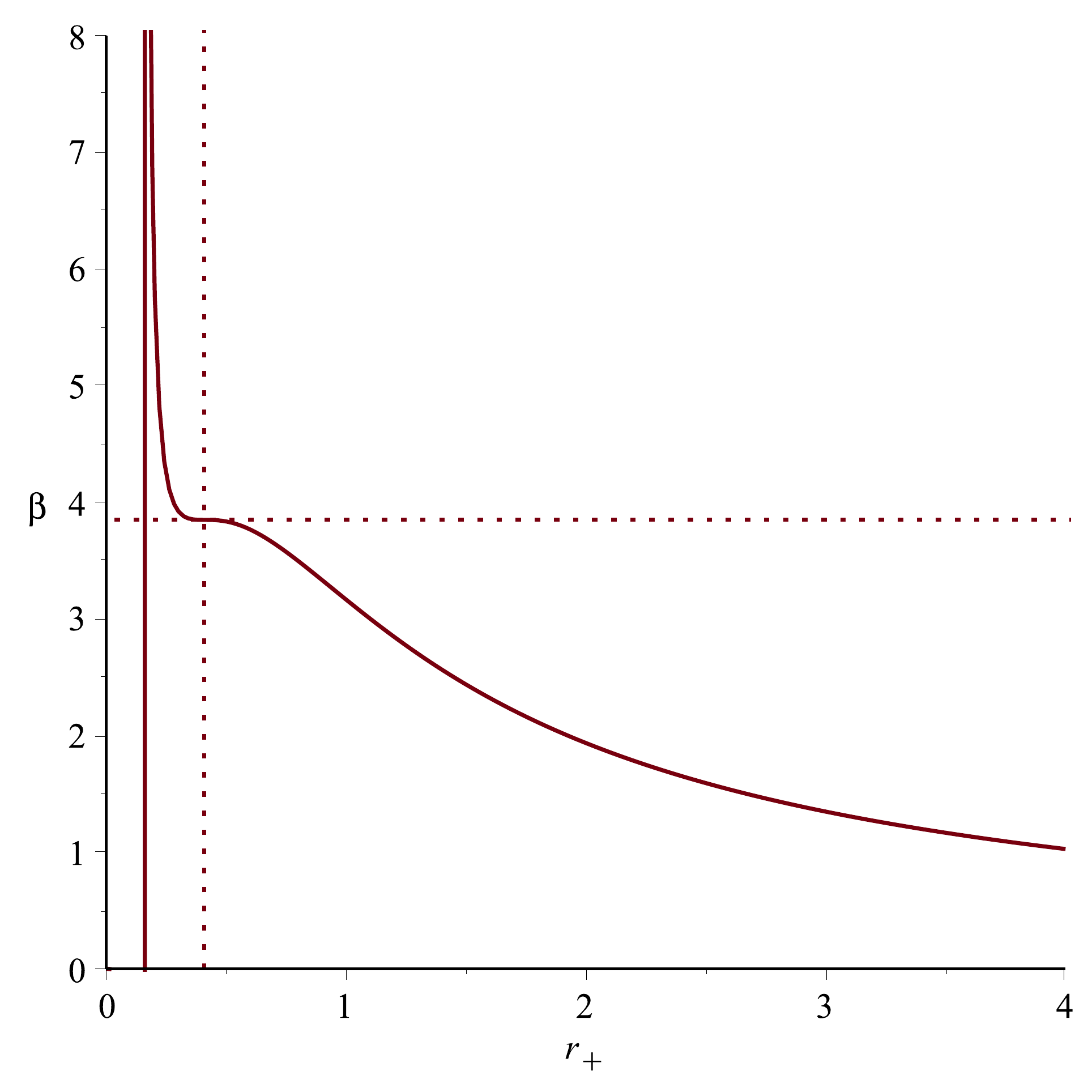}}
   \caption{\small Sample curves showing the shape of the inverse temperature as function of horizon radius for the fixed charge ensemble. The dotted line shows the location of the first order transition (on  the left) and a second order transition (on the right). See text for discussion.}  \label{fig:beta_curves2}
   \end{center}
\end{figure}
Below a certain temperature, there are no black holes and the only available solution is AdS. Eventually, the large and small (as measured by their relative horizon radii $r_+$) holes become available. The small branch are actually thermodynamically unstable (but interesting nonetheless, not the least because they are the ones that most resemble the Schwarzschild black holes of asymptotically flat spacetime), and in any case turn out to have action higher than that of the large black holes. The large black holes' action becomes lower than that of AdS at some temperature $T_*=\beta_*^{-1}$. For example for the case of $\Phi=q=0$ it is (in units where $l=1$) $\beta_*=\pi$.  See figure~\ref{fig:free_energy1}(a).

For $\Phi\geq1$ the structure is different (see figure~\ref{fig:beta_curves1}(b)). In that case, there are black holes for all temperatures. There is a minimum radius hole at $r_+=r_{\rm e}$ given by $3r_{\rm e}^4+l^2r_{\rm e}^2 -q^2l^2=0$, at which $T=0$. The ``small'' branch holes have disappeared, in essence, and all holes are large, and in fact there is now no transition at any temperature\footnote{Well, as far as this analysis is concerned. For this ensemble and the fixed charge ensemble, one can do further stability analysis to uncover more subtle phase structure. The extremal  black holes at zero temperature also have certain instabilities in this ensemble. This will not concern us here. See ref.\cite{Chamblin:1999hg} for further study of these issues.}. The relative free energy is negative for all~$\beta$. See figure~\ref{fig:free_energy1}(b).

Our interest in this paper is largely to map out how the entanglement entropy behaves in these finite volume systems, and to uncover features of its behaviour near the transitions. For our purposes, the first order transitions of the fixed potential ensemble are all somewhat analogous to each other in that they have a jump from pure AdS to the AdS black hole (with fixed potential).  We shall look at that in the next section.

\begin{figure}[h]
\begin{center}
\subfigure[$q<q_c=\frac{l}{6}$.]{\includegraphics[width=3.2in]{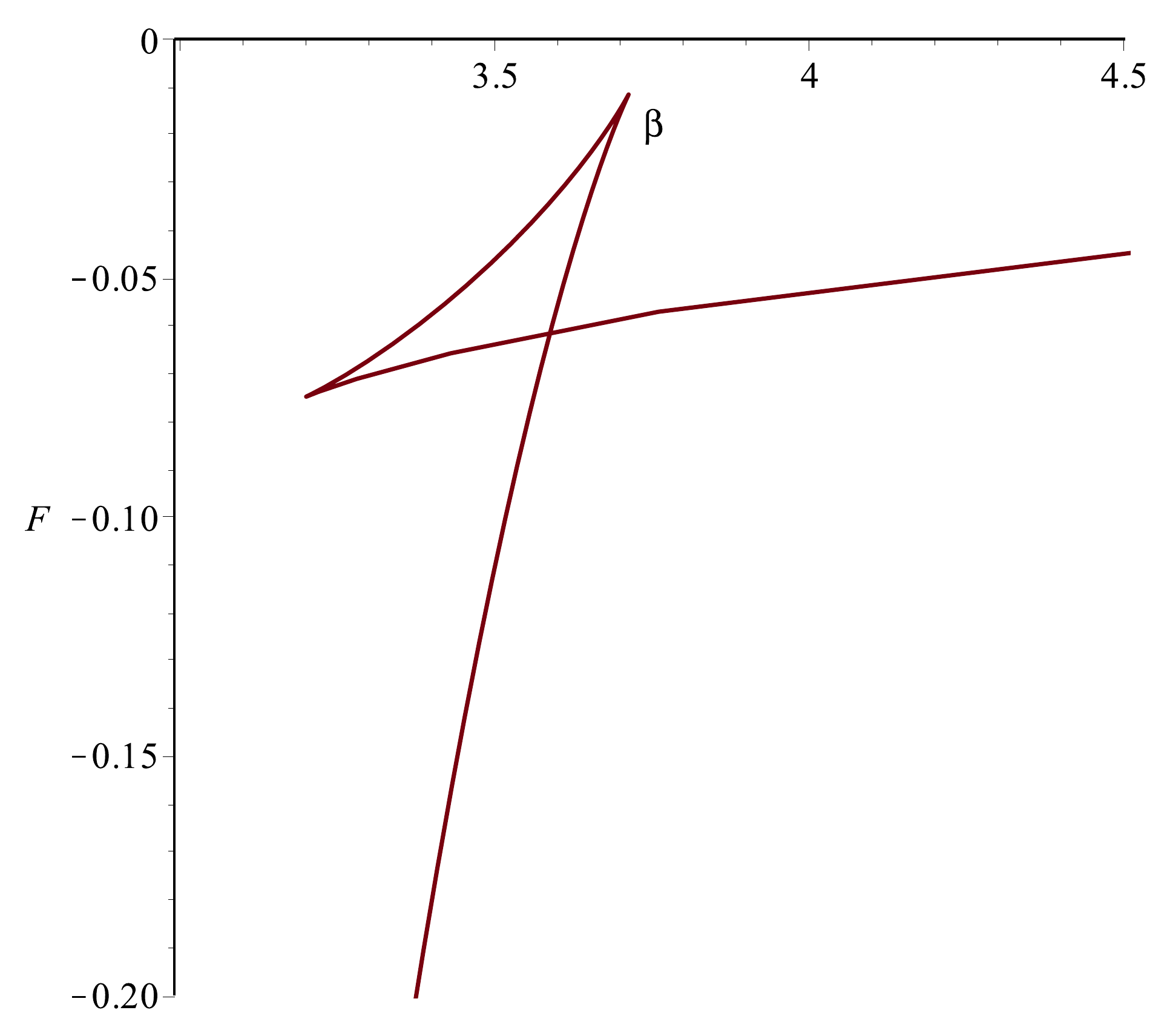}} 
\subfigure[$q=q_c=\frac{l}{6}$.]{\includegraphics[width=3.1in]{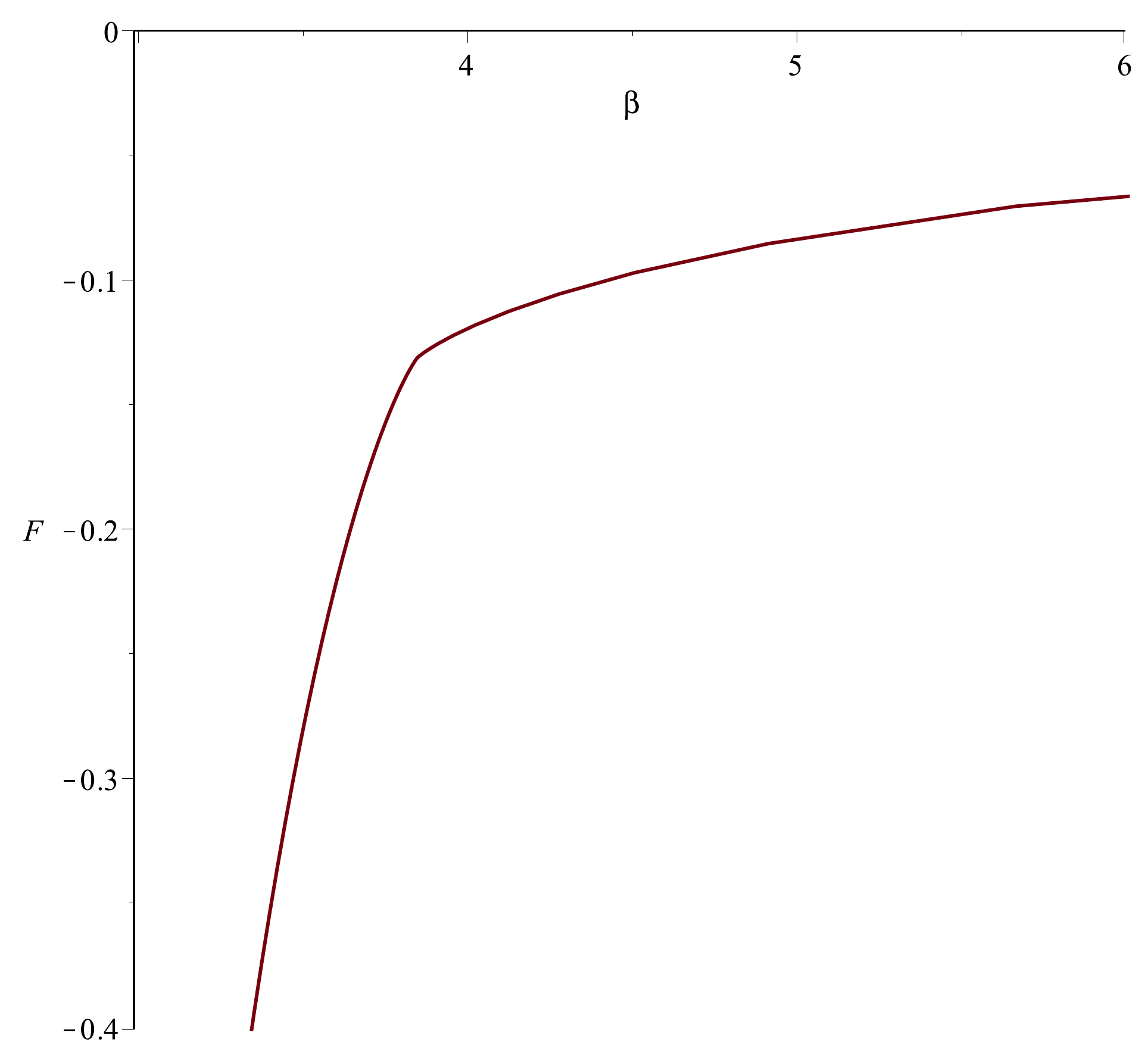}}
   \caption{\small The swallowtail and critical curves for free energy {\it vs.} $\beta$ in the fixed charge ensemble.}   \label{fig:free_energy2}
   \end{center}
\end{figure}

\begin{figure}[h]
\begin{center}
{\includegraphics[width=5.0in]{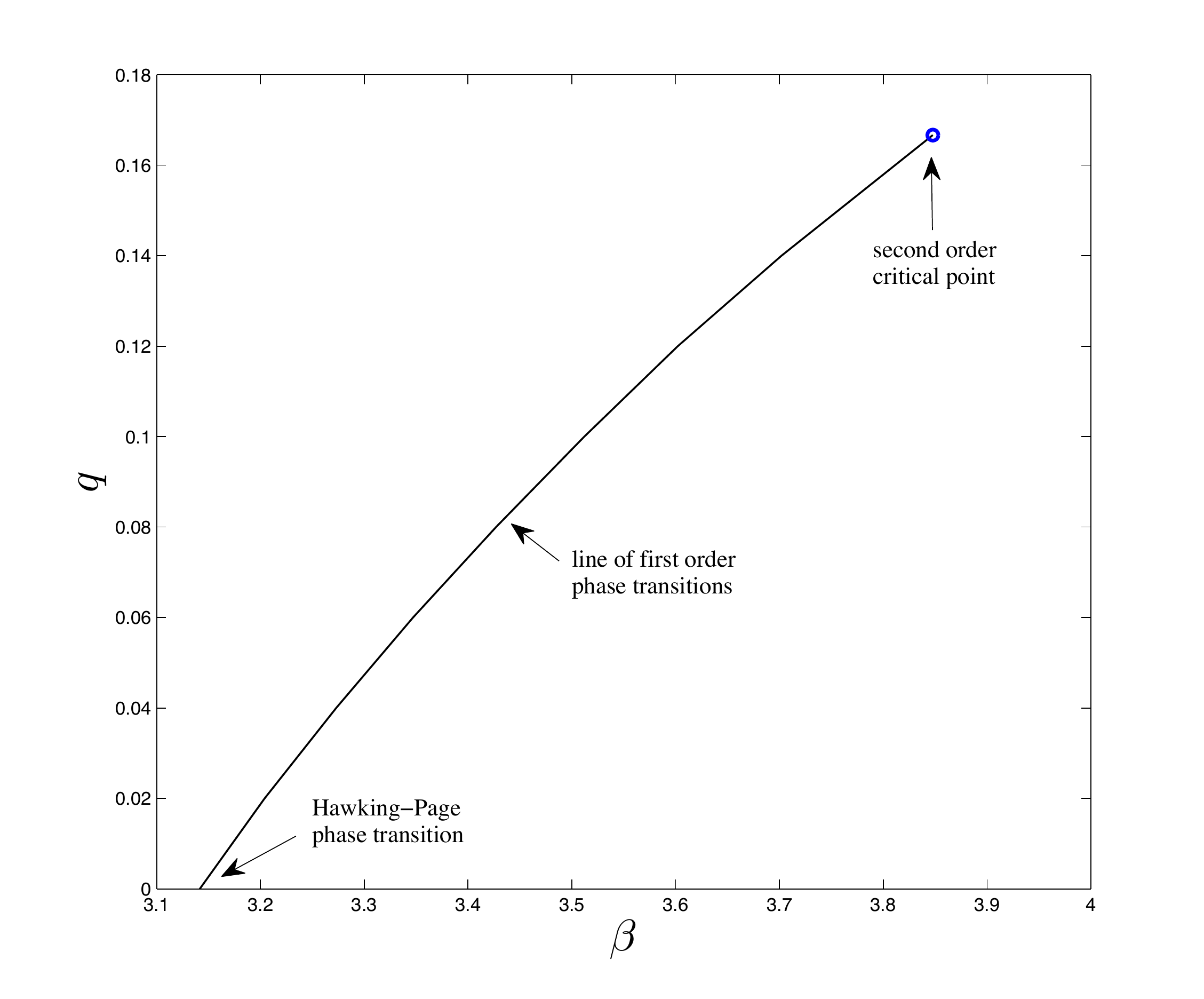}} 
   \caption{\small The phase diagram for the fixed charge ensemble. There is a line of first order points beginning with the Hawking--Page transition at $q=0$, $\beta=\pi$ and ending with a second order critical point at $q_c=1/6$, $\beta_c=\pi(3/2)^{1/2}$. Here, we used $l=1$ units. See text for discussion.}  \label{fig:phasediagram}
   \end{center}
\end{figure}

There are two intriguing features of the fixed charge ensemble that seem worth exploring. These were entirely new types of transition discovered in ref.\cite{Chamblin:1999tk}. They are transitions  between black holes of different sizes, and constitute a line of first order transitions that terminates in a second order point, in a manner analogous to the classic Van der Waals gas\footnote{The thermodynamics of these black holes can also be studied in an ensemble where the cosmological constant is allowed to vary, and in which the second order critical point   has an even stronger resemblance to the Van der Waals system. See refs.\cite{Kubiznak:2012wp,Gunasekaran:2012dq}.}. This is certainly an interesting family of transitions across which to track the entanglement entropy. Let us recall the structure. Plotting the ``equation of state'' $\beta(r_+)$ as a function of fixed $q$   shows that there are again two broadly different situations. For $q<l/6$, there are three families of black hole (see figure~\ref{fig:beta_curves2}(a)). The large (stable) and the small (unstable) branches of before are joined by  a new, even smaller branch which is in fact stable. There are black holes for all temperatures, starting out again at $T=0$ with the  extremal hole of radius $r_{\rm e}$, and going up from there. For a range of temperatures, there are then three holes competing thermodynamically, and for a while the smallest branch continues to win until at some $T_*$ the system jumps to the large holes. See the red dotted line in the figure. This is all made more clear by computing the relative free energy again (now using the fixed charge extremal black hole as the reference), which is\cite{Chamblin:1999tk}:
\begin{equation}
F=\frac{I}{\beta}=\frac{1}{4G l^2}\left(\l^2 r_+ - r_+^3+\frac{3q^2l^2}{r_+}-\frac{4}{3}l^2 r_{\rm e}-\frac{8}{3}\frac{q^2l^2}{r_{\rm e}}\right)\ .
\end{equation}
A plot of it reveals the beautiful swallowtail structure uncovered in ref.\cite{Chamblin:1999tk}, showing how the three branches fit together. See  figure~\ref{fig:free_energy2}(a).
The middle (unstable) branch of black holes makes the base of the tail, and the transition from the smallest  to the large branch is along the bottom edge of the structure, at the kink (running in from the right to increase the temperature, for example). The first order transitions that occur at this kink go away when the we get to $q=q_c=l/6$, since the tail shrinks away to leave a smooth structure, the result of the smallest and largest branches merging into one (see figure~\ref{fig:beta_curves2}(b) and figure~\ref{fig:free_energy2}(b)), and squeezing out the unstable branch. Intriguing is the second order point at which this happens (see  figure~\ref{fig:phasediagram} for a phase diagram). This is perhaps the earliest example in the literature of a second order point for a gauge theory realized as a gravity dual, and since there is a lot of  current interest in such points  (for example in holographic studies of superconductivity\cite{Gubser:2008px,Hartnoll:2008kx} and related novel quantum phases of potential experimental interest\cite{Herzog:2009xv,Hartnoll:2009sz}), it is worth exploring how the entanglement entropy behaves near it, perhaps as a model for other systems (even though it is of a somewhat different sort from the systems currently being studied).
%
\section{Entanglement Entropy in Global Coordinates}
\label{sec:eeglobal}
The entanglement entropy for systems asymptotic to global AdS does not seem to have been studied in the literature as much as for local AdS, and so we shall go carefully, so as to be clear.
The spatial part of the  field theory in our case is on a round $S^2$, with coordinates $\theta$ and $\varphi$. For the computations of entanglement entropy, geometrically, region ${\cal A}$ will be parameterized by a shape in~$(\theta,\varphi)$. Our choice for the boundary will either be constant $\theta$, giving a cap/yarmulke or disc shape, or constant~$\varphi$, giving what begs to be called an orange slice\footnote{Transitions in entanglement entropy and geometrical entropy were studied in refs.\cite{Faraggi:2007fu} and \cite{Bah:2008cj} in fixed black hole geometries, by comparing minimal surfaces to surfaces that end on the horizon, as a function of the size of  region~${\cal A}$. Our goals in this paper are different, since we will be fixing the size of region~${\cal A}$  and instead allowing  the bulk geometry to make transitions.}. In a sense, these are the natural finite volume analogues of the disc and strip that are commonly studied in the literature for infinite volume.   A key additional difference here is that due to our choices, the {\it finite} complementary region,~${\cal B}$ is  in fact of the same shape as ${\cal A}$. In fact there is a point where  ${\cal A}$  begins to become larger than~${\cal B}$, and by symmetry, it is clear that the form of the entropy will be the same as it was for smaller ${\cal A}$,  regardless of what is happening in the bulk. There'll be an explicit analytic example below for pure AdS$_4$. When there is non--trivial topology in the bulk due to a black hole horizon, there  is  an important caveat to the previous observation. A refinement of the Ryu--Takayanagi prescription\footnote{See {\it e.g.,} refs. \cite{Ryu:2006ef,Fursaev:2006ih,Headrick:2007km,Emparan:2006ni,Blanco:2013joa} for more discussion. Thanks to R. C. Myers and M. Rangamani for useful communications about this.} shows that for large enough ${\cal A}$, the entropy can indeed be written in terms of the minimal surface associated with ${\cal B}$, {\it plus} a contribution from the surface that wraps the horizon. For the studies in all of this paper, we will stay away from this (strongly thermal) regime for the entropy. 

The  profile in the bulk will be given by the function $r(\theta,\varphi)$. The induced metric on the surface, with coordinates $(\xi^a, a=1,2)$ is:
\begin{equation}
h_{ab}= G_{\mu\nu}\frac{\partial x^\mu}{\partial \xi^a} \frac{\partial x^\nu}{\partial \xi^b}\ ,
\end{equation}
where we pick $\xi^1=\theta$ and $\xi^2=\varphi$. So the area of our surface comes from minimizing the following problem:
\begin{equation}
A=\int_{\theta=0}^\pi \!\int_{\varphi=0}^{2\pi} \!\! ({\rm det } h)^\frac12\, d\theta d\varphi\ .
\end{equation}
Since our only non--trivial embedding function  $x^\mu(\xi^a)$ is $r(\varphi,\theta)$,  the Lagrangian is a classical mechanics problem with either $\varphi$ or $\theta$ acting as a time parameter:
\begin{equation}
{\cal L}=\left[G_{\varphi\varphi}\left(G_{\theta\theta}/2+G_{rr}(\dot r)^2\right)+G_{\theta\theta}\left(G_{\varphi\varphi}/2+G_{rr}(r^\prime)^2\right)\right]^{\frac12}\ ,
\end{equation}
where ${\dot r}\equiv \partial r/\partial\theta$ and $r^\prime\equiv \partial r/\partial\varphi$.

\subsection{Orange Slices}
Here, pick a surface bounded by the two lines of longitude $\varphi=\pm\varphi_0$.  Then the problem of  minimization can be treated by having $\varphi$ act as a time, with the benefit that  since the metric coordinates $G_{\mu\nu}$ have no $\varphi$ dependence, the Hamiltonian derived from the Lagrangian is conserved. The Hamiltonian's value can be determined by noting that at the point $(\varphi=0,\theta=\pi/2),$ both $r^\prime$ and $\dot r$ vanish, and $r$ attains  its maximum value, $r_0$. However, the $\theta$ dependence of the Lagrangian means that even with this result, the problem remains a difficult one to solve\footnote{The earlier versions of this manuscript presented an attempt to reduce the problem to a one dimensional first order problem. The accompanying ansatz and reduced Lagrangian  were incorrect. We thank M. Rangamani for pointing out the error.}, and we shall not pursue it here. So unlike  the strip, its cousin in local coordinates, the orange slice geometry does not drastically simplify, and it is the disc that shall turn out to be the easier case to study,  as we shall see next.

\subsection{Discs}
\label{sec:disc}
In this case consider a surface bounded by the line of latitude $\theta=\theta_0$, which  has the topology of a disc. Then $\theta$ can be treated as a time. Because of the lack of $\varphi$ dependence of the metric components, and hence of the Lagrangian, every $\varphi$ is equivalent, and hence we can drop the $\varphi$ dependence and treat the problem of solving for $r(\theta)$ as that of a particle. Then the area is:
\begin{equation}
A= 2\pi \int_0^{\theta_0}G_{\varphi\varphi}^{\frac12} \left(G_{\theta\theta}+G_{rr}(\dot r)^2\right)^{\frac12} \ ,
\end{equation}
where here ${\dot r}=dr/d\theta$. The (rather complicated) equation of motion for $r(\theta)$ must be solved with the boundary condition ${\dot r}=0, r=r_0$ at $\theta=0$, and $r=\infty $ at $\theta = \theta_0$. We regulate the area by integrating out not to $r=\infty$ but $r=l^2/\epsilon$, for small $\epsilon$.
In general this problem is difficult  to solve, and an analytic expression does not seem accessible. There are no nice conservation laws to readily give us a first integral. Even in the pure (global) case of $q=m=0$ it is a rather difficult task to solve directly, with the equation of motion being:
\begin{eqnarray}
&& \sin  \theta\left( {l}^{4} r ^{2}+ r ^{4}{l}^{2}
 \right)  {\ddot r} +{l}^{4}\cos \theta ({\dot r})^{3}- \sin\theta  \left( 4\,
r^{3}{l}^{2}+3\,r {l}^{4} \right)  ({\dot r})^{2} \\ &&\hskip4cm  + \cos \theta\left( {l}^{4} r ^{2}+ r ^{4}{l}^{2} \right) 
 {\dot r} -  \sin \theta \left( 2\,r ^{7}+4\,{l}^{2} r ^{5}+2\, r ^{3}{l}^{4} \right)
 =0\ . \nonumber
\end{eqnarray}
In fact, some experimentation shows that there is an exact solution to this equation, which is\footnote{Actually, in retrospect it can be readily deduced from various  changes of variable to be found in ref.\cite{Casini:2011kv} (we thank Tameem Albash for a reminder to carefully read that paper), or in refs.\cite{Hubeny:2007xt,Hubeny:2012wa}.}:
\begin{equation}\label{eqn:exact}
r(\theta)=l\left[\left(\frac{\cos\theta}{\cos\theta_0}\right)^2-1\right]^{-\frac12}\ ,
\end{equation}
with 
\begin{equation}
\cos\theta_0=\frac{r_0}{\sqrt{l^2+r_0^2}}\ .
\end{equation}
The area integral can be evaluated with the result:
\begin{equation}
\label{eq:pureadsresult}
A^\circ= 2\pi l^2\left[\frac{l}{\epsilon}\left(1+\frac{\epsilon^2}{l^2}\right)^{\frac12}\sin\theta_0-1\right] \simeq 2\pi l^2\left[\frac{l}{\epsilon}\sin\theta_0-1\right] \ .
\end{equation}
This is the expected form, with the area law coming with the UV cutoff, where the ``area" of the boundary of the region in question is the perimeter of the disc of radius $l\sin\theta_0$, and with the appropriate universal term. This behavior is typical of what we'll see later as well. There's a maximum value that the entanglement entropy $S(\theta_0)=A/(4G)$ can attain, quite naturally since at $\theta=\pi/2$ the system starts shrinking again, by symmetry, a consequence of finite volume. The $\theta_0\to\pi-\theta_0$ symmetry is manifest.

To make further progress one can delve into the delicate matter of extracting numerical results for $m$ and $q$ non--zero. One must solve the second order equation of motion for $r(\theta)$ with the boundary condition, and then put $r(\theta)$ and ${\dot r}(\theta)$ into the area integral and integrate up to the cutoff. It is actually quite tractable, with some care, and it is worth recording one way of proceeding here. First, it it natural to treat the system as  a boundary value problem. Such problems, as numerical endeavours, benefit from a good starting mesh from which to iterate to convergence. The exact solution~(\ref{eqn:exact}) is  very useful to have to input as a starting mesh. MatLab's {\tt bvp4c} and {\tt bvp5c} were used here, with tight error tolerances switched on, and grids  of sizes ranging from $10^4$ to $10^6$, depending upon where in parameter space was being probed. We worked with $G=1$,  $l=1$, and $\epsilon=10^{-4}$. It is important to respect staying on the solution of interest while implementing   the cutoff of the integration (and the finiteness necessary to do the boundary value problem) by appropriate tuning of the boundary conditions in the numerical problem. 

The numerical plots for $S(\theta_0)$ for fixed generic $m$ and $q$ are not particularly illuminating. The area law is readily verified (in a wide range of regimes where the numerics are reliable) and is of the same form as shown above in equation~(\ref{eq:pureadsresult}), since, of course, the boundary geometry is the same. That contribution dominates, and the subleading physics is the $m,q$ dependence that is present for a given $\theta_0$. We see, for example, the entropy rise with temperature (we will be more explicit below). We can  subtract off the entanglement entropy of the pure AdS part, to display more directly the  interesting physics.   This will be done in the next section. For our purposes, the exact $\theta_0$ dependence is perhaps not so interesting to display in a series of curves. So we will not. What remains to be done is our main task of looking at the entropy's response to the first and second order phase transitions we reviewed earlier.

\section{Entanglement Entropy and Large $N$ Phase Transitions}

The disc regions of section~\ref{sec:disc} can be readily made to yield a wealth of information about the entanglement entropy $S(\theta_0)$ in the neighborhood of the phase transitions.  It is important to note that here we are not considering transitions in the entropy  as a function of, say, $\theta_0$,  for a fixed background (see footnote~4), nor as a function  of temperature. The fact that the {\it backgrounds} make transitions is something the  gravity theory has established for us already (and which we reviewed in section~3). So we would like to follow the entropy by tracking it from one background to another. It is a probe of the bulk geometry, not a driver in and of itself. In this way we hope to learn from these models more about how it may be used as a diagnostic tool in other systems. One way to proceed with uncovering our physics is to stay at a fixed $\theta_0$ (the choice $\theta_0=0.005$ was made) and  scan the space $(m,q)$ of solutions, whether they be small or large (or very small) black holes, computing the entanglement entropy for them. Then folding in the equation of state information {\it i.e.,} $\beta(r_+)$, one can arrive at the entropy  as a function of $\beta$ for the various branches. Knowledge of the branches that are selected at various points in the $(T,\Phi)$  or $(T,q)$ plane shows how to move from one entropy curve to another, either smoothly or discontinuously, as dictated by the bulk thermodynamics described in section~3.

\begin{figure}[h]
\begin{center}
{\includegraphics[width=5.0in]{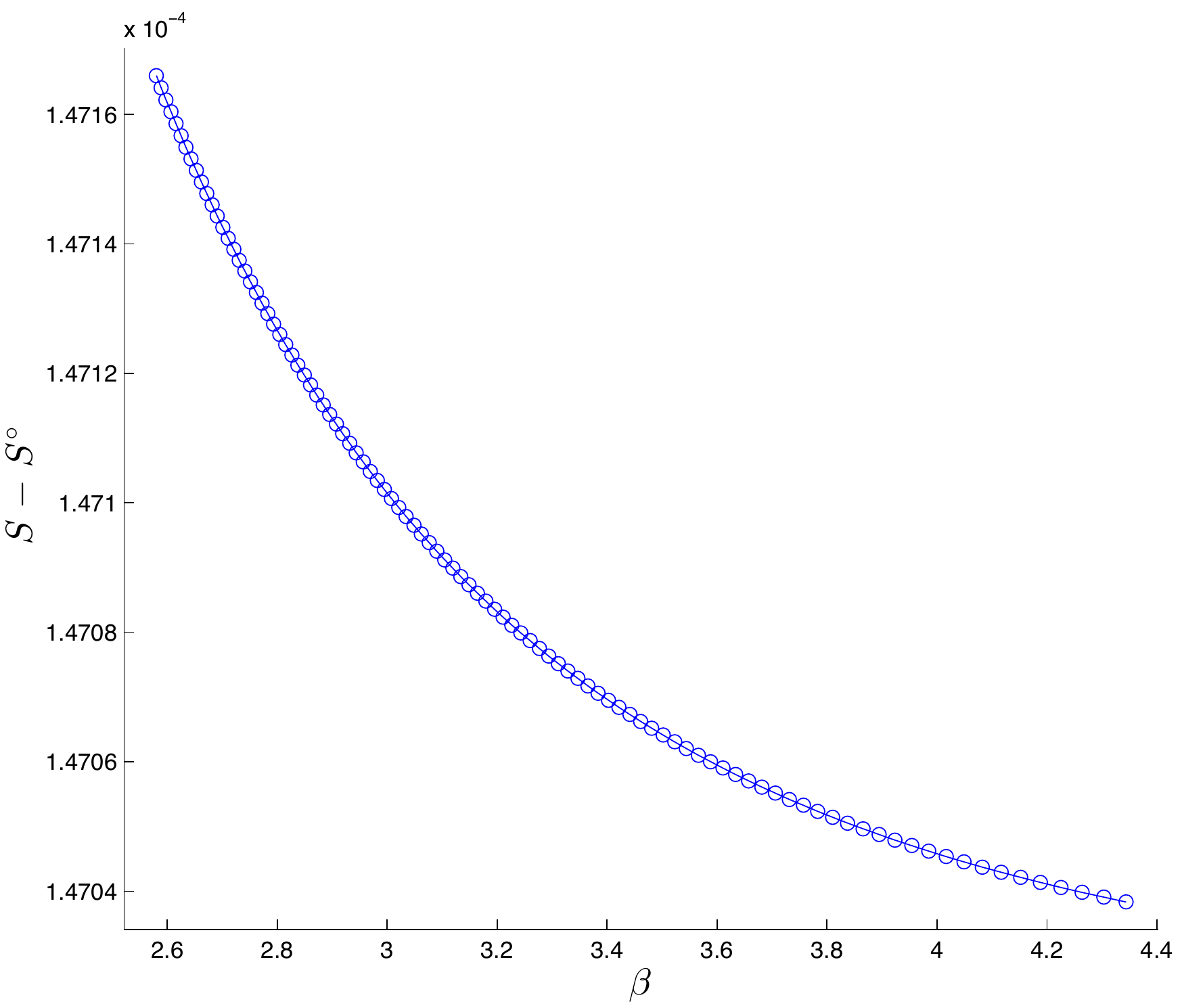}} 
   \caption{\small The disc entanglement entropy for fixed angle {\it vs} inverse temperature for black holes in the fixed charge ensemble for $q>q_c$.  The same qualitative behavior is seen for the fixed potential ensemble with~$\Phi>\Phi_c$. The quantity $S^\circ\simeq 76.96869316$ is the pure AdS contribution.  See text for discussion.}  \label{fig:ee_vs_beta1}
   \end{center}
\end{figure}

\begin{figure}[h]
\begin{center}
{\includegraphics[width=5.0in]{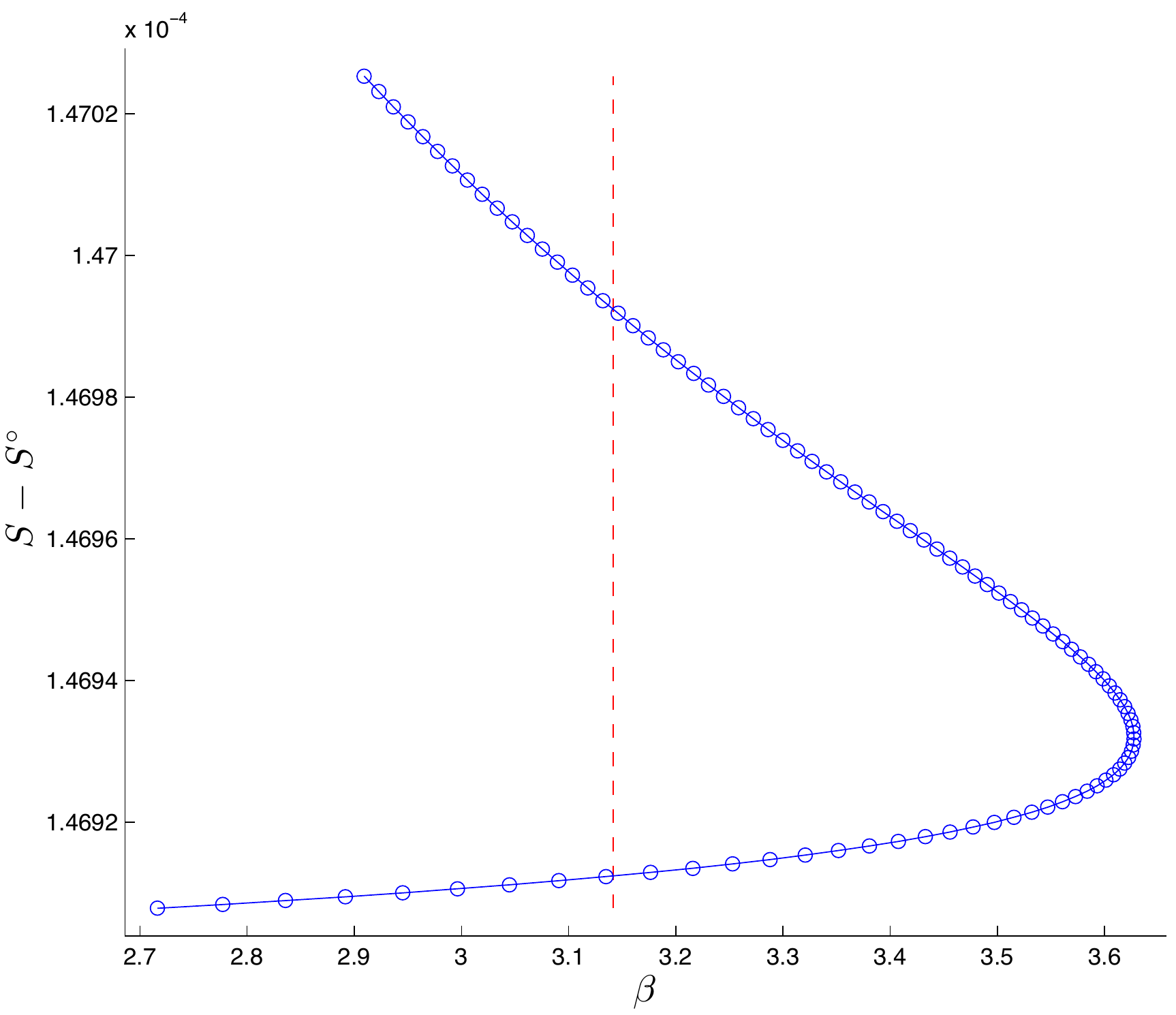}} 
   \caption{\small The disc entanglement entropy for fixed angle {\it vs} inverse temperature for black holes in the fixed potential ensemble with~$\Phi<\Phi_c$. The red dotted line marks the transition temperature where AdS hands over to the black holes for  high enough temperature. Then the upper branch is the relevant entropy.  The quantity $S^\circ\simeq 76.96869316$ is the pure AdS contribution.  See text for discussion. }  \label{fig:ee_vs_beta2}
   \end{center}
\end{figure}

The simplest case is of course when there is no transition, with large black holes the favoured phase for all temperatures. This is either in the fixed~$\Phi$ ensemble for  $\Phi>\Phi_c=1$, or the fixed charge ensemble for $q>q_c=l/6$. In such cases, we have the situation shown in figure~\ref{fig:ee_vs_beta1}, where the choice $G=1$, $l=1$, and $q=1$ was made. In the figure, we have subtracted the overall    pure AdS contribution $S^\circ=A^\circ/4=(\pi/2)((1+\epsilon^{-2})^{1/2}\sin\theta_0-1)\simeq76.96869316$  seen in the last section (see equation~(\ref{eq:pureadsresult})). There we see that the entropy  smoothly varies as a function of temperature (rising with it). 
%
(This plot was done for the fixed charge ensemble at $q=1$, but  also qualitatively illustrates the key features of  the fixed $\Phi$ ensemble for $\Phi>\Phi_c=1$.) The next case is the fixed~$\Phi$ ensemble  for $\Phi<1$, and we chose parameters $G=1, l=1, q=\Phi=0$ for illustration. See figure~\ref{fig:ee_vs_beta2}.  Again $S^\circ$ is the pure AdS contribution that has been subtracted off. As the temperature is raised (coming in from the right), the favoured entanglement entropy is that of pure AdS, since it is thermodynamically preferred, until a transition at~$\beta_*$ (which is  exactly $\pi$ for this case). 
 At that point, the plotted curve becomes relevant, and the entropy jumps discontinuously from that of AdS to the (higher) value that the large (upper) branch holes have (the upper part of the curve -- the lower branch plays no role here), and then  proceeds along on its rise with temperature\footnote{Note that there is no curve showing the temperature dependence of the entropy for pure AdS. There is presumably such a dependence for the dual field theory (the confined phase), but the holographic prescription does not capture it since the geometry sees the temperature in the periodicity of $\tau$, the analytically continued Euclidean time, and the minimal surface is at fixed time. It is clear, nonetheless, that it starts out less than that of the black hole entanglement entropy since there the number of degrees of freedom are known to go from being order one to order~$N^2$ when the black hole is favoured (now being the dual to the unconfined phase\cite{Witten:1998zw}). This is  responsible  for the entropy's jump.}.

\begin{figure}[h]
\begin{center}
{\includegraphics[width=5.0in]{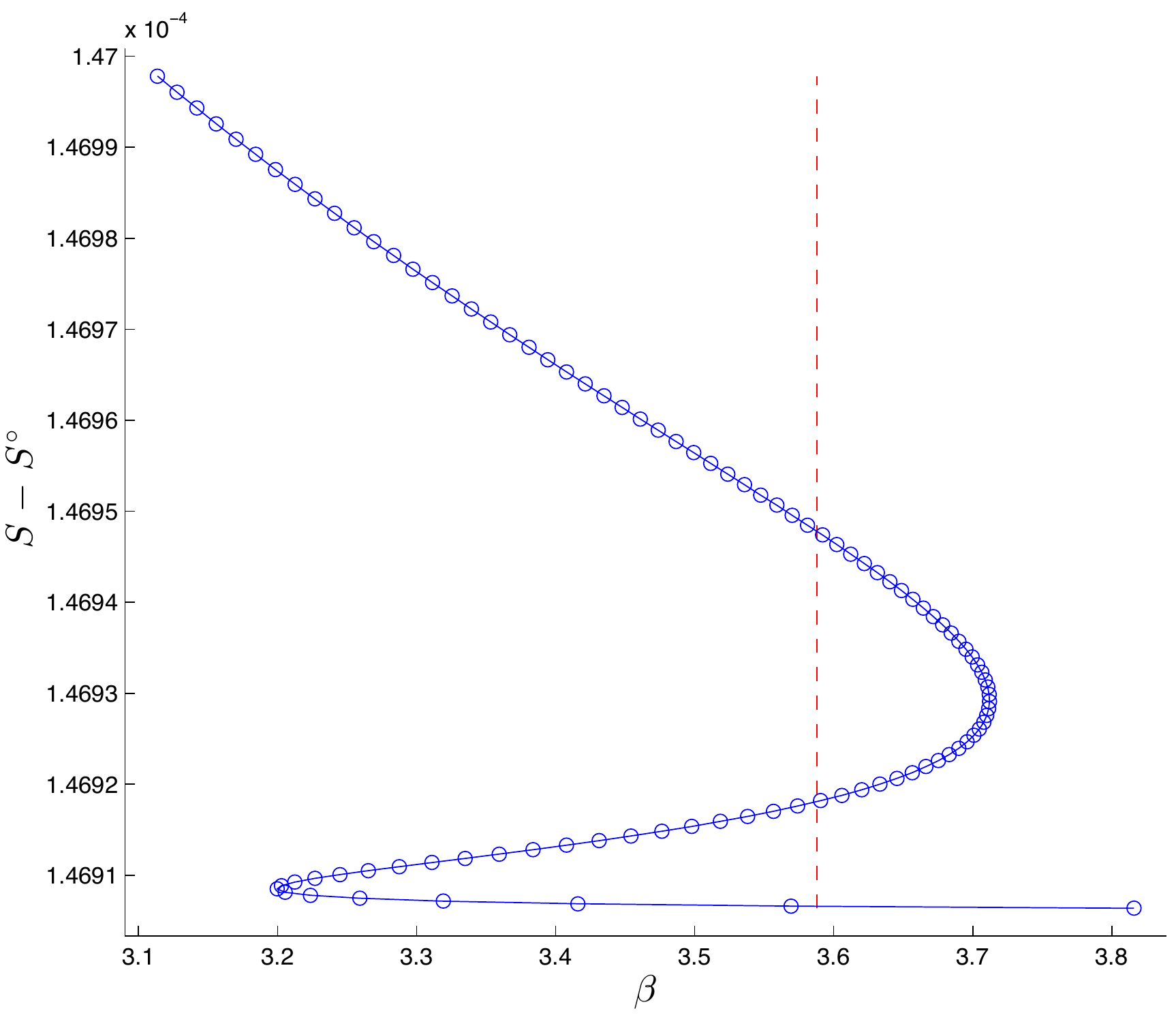}} 
   \caption{\small The disc entanglement entropy for fixed angle {\it vs} inverse temperature  for black holes in the fixed charge ensemble with~$q<q_c$. The red dotted line marks the transition temperature where the smallest family of black holes (lowest branch) hands over to the largest family (upper branch) at high enough temperature. The quantity $S^\circ\simeq 76.96869316 $ is the pure AdS contribution.  See text for discussion. }  \label{fig:ee_vs_beta3}
   \end{center}
\end{figure}

The fixed charge ensemble's black hole to black hole transitions are next. See figure~\ref{fig:ee_vs_beta3}.
 (The parameter values $G=1, l=1, q=1/6 - 0.05$ were chosen, and again $S^\circ$  is the pure AdS value.) There we see all three branches' entanglement entropy.  Along the bottom of the figure coming in from the right (rising temperature) is a slowly rising  entanglement entropy, along the lowest branch. Again the dotted (red) line shows the transition temperature $\beta_*$, which is $\simeq 3.588$ for this example. At the transition, there is a discontinuous (albeit modest) jump in the entropy to the upper branch (the large black holes) where henceforth it rises more rapidly with temperature. 

\begin{figure}[h]
\begin{center}
{\includegraphics[width=5.0in]{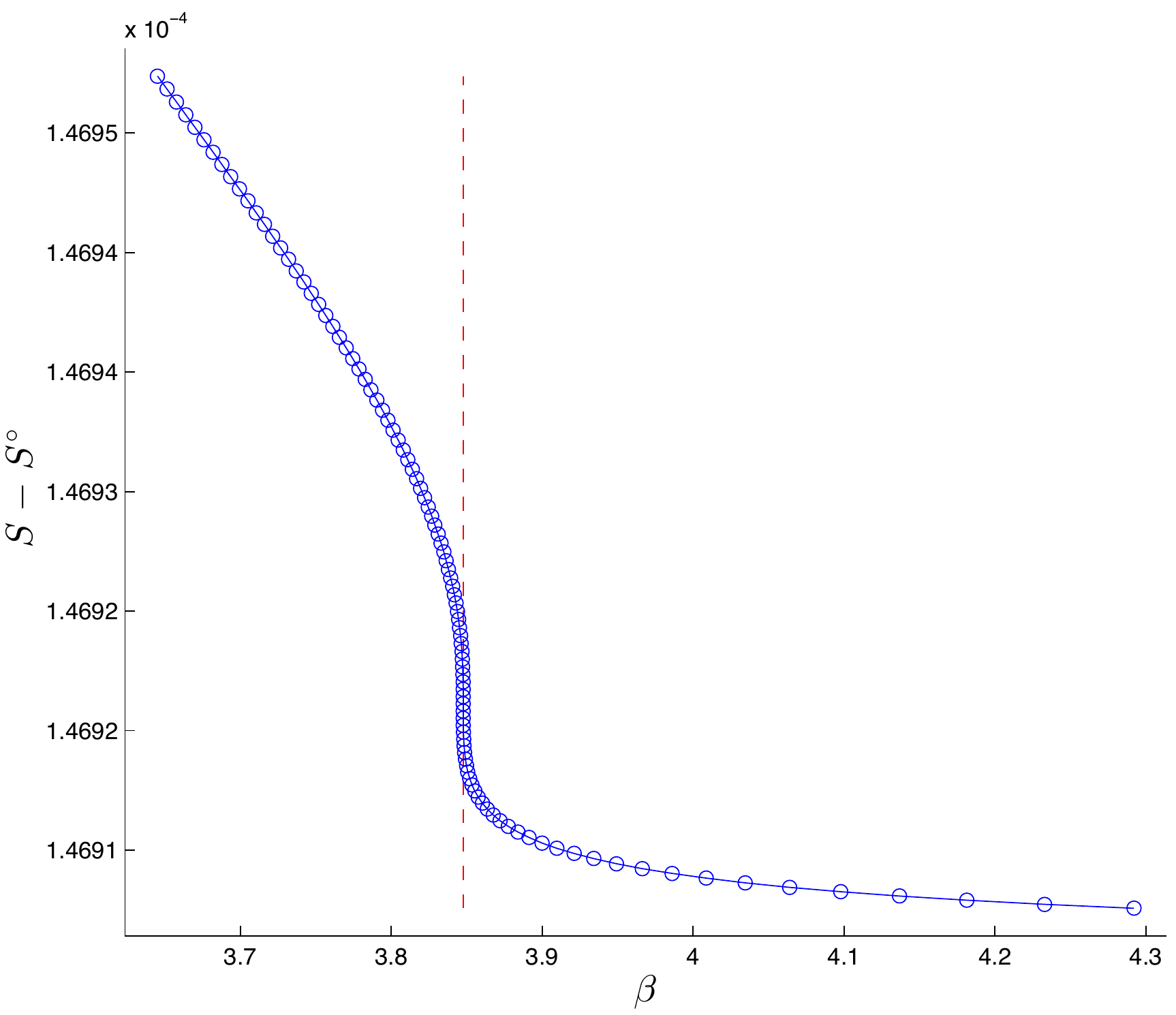}} 
   \caption{\small The disc entanglement entropy for fixed angle {\it vs} inverse temperature  for black holes in the fixed charge ensemble with~$q=q_c=1/6$. This is a second oder critical point. The red dotted line marks the transition temperature at $\beta_c=\pi (3/2)^\frac12$. The quantity $S^\circ\simeq 76.96869316 $ is the pure AdS contribution.  See text for discussion.}  \label{fig:ee_vs_beta4}
   \end{center}
\end{figure}

Finally, we turn to the second order transition and its neighbourhood. This is shown in figure~\ref{fig:ee_vs_beta4}, where $G=l=1$ and $q=q_c=1/6$, with $S^\circ$  the pure AdS value.  We are rewarded with a remarkable shape. 
The modest rise with temperature on the lower branch is smoothly connected to the more rapid rise with temperature {\it via} a stationary point, at $\beta_c=\pi (3/2)^\frac12$. 

\begin{figure}[h]
\begin{center}
\subfigure[$q<q_c=\frac{l}{6}$.]{\includegraphics[width=3.2in]{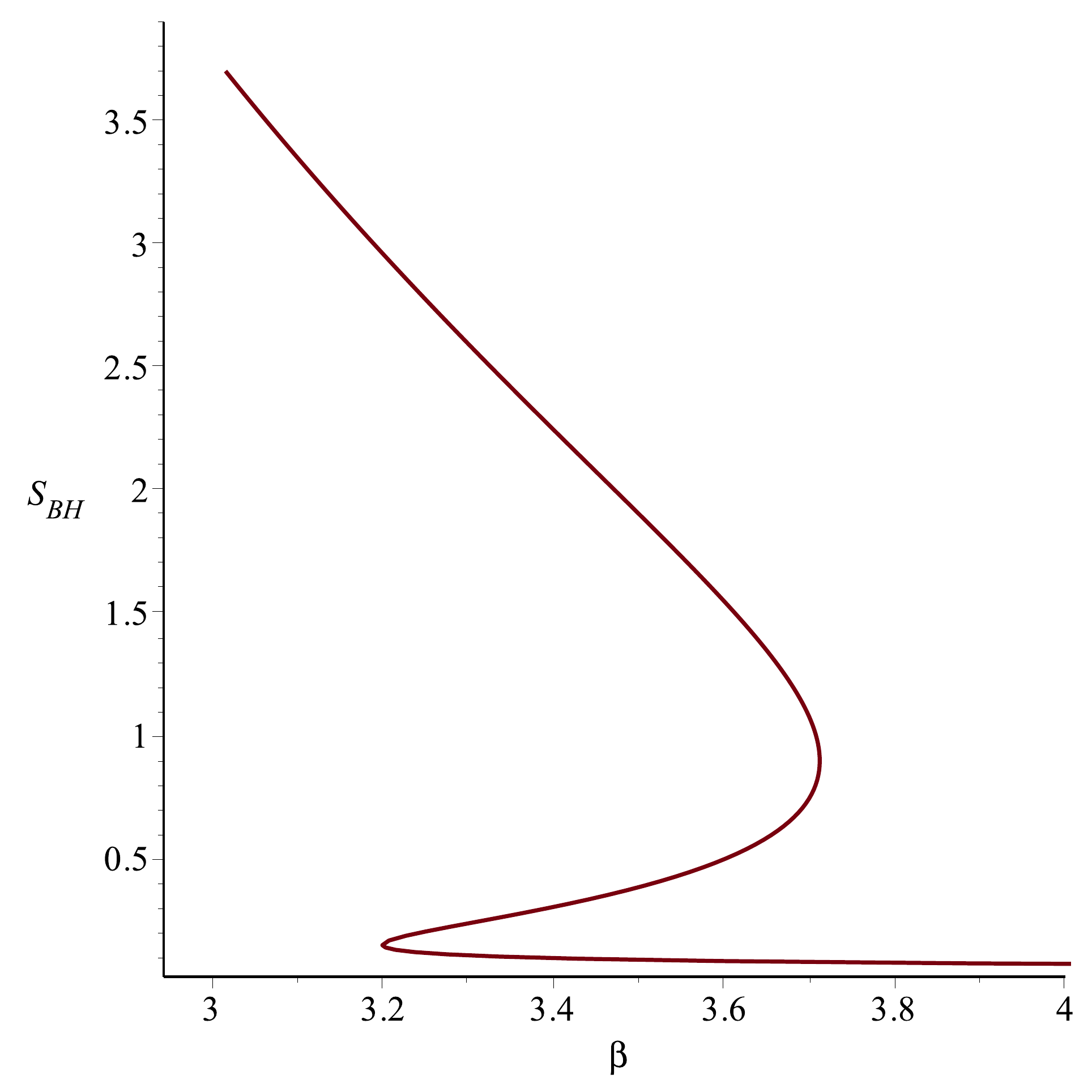}} 
\subfigure[$q=q_c=\frac{l}{6}$.]{\includegraphics[width=3.1in]{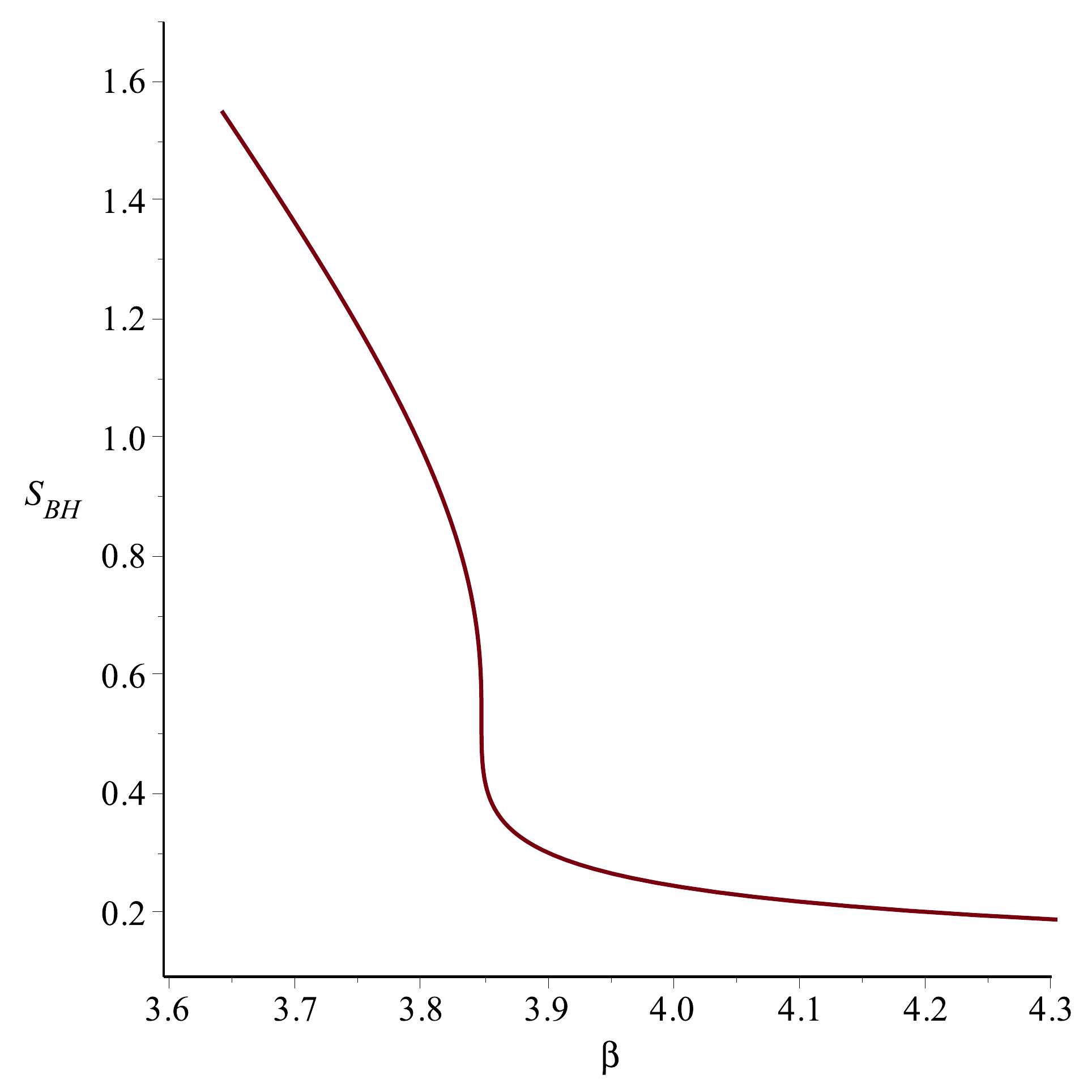}}
   \caption{\small The black hole entropy {\it vs.} $\beta$ for subcritical ($q=l(1/6-0.05)$) and critical ($q=l/6$) charges. We used $l=1$ in the plots. Compare to the entanglement entropy curves given in figure~\ref{fig:ee_vs_beta3} and figure~\ref{fig:ee_vs_beta4} respectively, for the same choices of values of $q$.}   \label{fig:entropy}
   \end{center}
\end{figure}

It is interesting to notice that some of the features of the entanglement entropy that we've uncovered are actually familiar. For the fixed charge ensemble, we can see that the curves in figures~\ref{fig:ee_vs_beta3} and~\ref{fig:ee_vs_beta4} are somewhat reminiscent of the $(\beta,r_+)$ curves in figure~\ref{fig:beta_curves2}, while for fixed potential, the curves in figures~\ref{fig:ee_vs_beta1} and~\ref{fig:ee_vs_beta2} resemble those for  $(\beta,r_+)$ in figure~\ref{fig:beta_curves1} . This is not an accident, and we can  unpack this somewhat. The Bekenstien--Hawking entropy\cite{Bekenstein:1973ur,Bekenstein:1974ax,Hawking:1974sw,Hawking:1976de} for any of our black holes, with horizon radius $r_+$, is $S_{BH}=\frac{1}{4}A_{BH}=\pi r_+^2$, in units where $G=1$. Our ``equation of state" in equation~(\ref{eq:equationofstate}) can be rewritten as an exact relation between the entropy and the temperature by substitution:
\begin{equation}
\beta=\frac{4\sqrt{\pi S_{BH}}}{3S_{BH}/{\pi l^2}+1-\Phi^2 }\  ,
\quad {\rm or} \quad
\beta=\frac{4\sqrt{\pi S_{BH}}}{3S_{BH}/{\pi l^2}+1-{q^2\pi}/{S_{BH}}}\  ,
\label{eq:betaess}
\end{equation}
for the fixed potential and fixed charge ensembles respectively. Let
us focus on the  fixed charge ensemble,  although the key remarks to be made about the similarity in shape apply to the fixed potential ensemble equally well.  Notice that for real $\beta$, $S_{BH}$ must be positive. In this  range, there are two branches. $\beta$ diverges at the positive root of the denominator, at
\begin{equation}
S_{BH}^e=\frac{l^2\pi}{6}\left[\left(1+\frac{12\pi q^2}{l^2}\right)^{1/2}-1\right] \ .
\end{equation}
(This is in fact the entropy of the extremal black holes, for which $T=0$.)
For $S_{BH}$ above that root,~$\beta$ is positive, while below it is negative. The turning points of $\beta(S_{BH})$  are located at the roots of 
\begin{equation}
{3S_{BH}^2}/{\pi l^2}-S_{BH}+{3q^2\pi} =0\ ,
\label{eq:criticalcondition}
\end{equation}
at
\begin{equation}
S_{BH}^\pm=\frac{l^2\pi}{6}\left[1\pm\left(1-\frac{36\pi q^2}{l^2}\right)^{1/2}\right] \ ,
\end{equation}
coalescing at the critical point $q=q_c=l/6$. The physical quadrant of the function $\beta(S_{\rm BH})$ is plotted in figure~\ref{fig:entropy} for the two sample values of $q$  we used before  ($q_c$ and $q_c-0.05l$), showing the similarity to our numerically obtained  curves for  the entanglement entropy.

Now all this is relevant to the entanglement entropy curves since, after  subtracting off the temperature independent contribution $S^\circ$ as we have done, in the regime when  region~${\cal A}$ is large enough ({\it i.e.,} $\theta_0$ is large enough), the entropy will see the thermal contribution, given by the black hole entropy  $S_{BH}$ since, as already mentioned near the start of section~4, there will be a contribution from wrapping the entire horizon.  So in that strongly thermal regime our curves should indeed coincide. However, we are far from this thermal regime, since we are working  at small $\theta_0$,  but interestingly we see that  the temperature dependence of the entanglement  entropy $S(\beta)-S^\circ$ resembles a  scaled cousin of the function $S_{BH}(\beta)$. By construction, the  function $S(\beta)$  naturally inherits multiple branches, like $S_{BH}(\beta)$ has (upper, lower and middle for  $q<q_c=l/6$), and it also has a  special point  at $q_c=l/6$ where  the middle branch disappears as the upper and lower branches merge. This alone does not imply that the  functional dependence of $S(\beta)-S^\circ$ is exactly the same as for $S_{\rm BH}(\beta)$, up to an overall scaling\footnote{In fact one can  scale, {\it e.g.},  the function in figure~\ref{fig:ee_vs_beta3} and see that it cannot match that in figure~\ref{fig:entropy}(a) everywhere.}. Crucially, the neighbourhood of the point where the branches merge may be a critical point of a different type than that seen for the black hole physics, only coinciding with it in the thermal regime. That the functions have the same qualitative  structure {\it does} allow us, by reference to $S_{\rm BH}(\beta)$,  to phrase some specific questions about the form  of~$S(\beta)$ in the neighbourhood of the  critical point, which we then can explore numerically.

\begin{figure}[h]
\begin{center}
\subfigure[Fit: $C(T)\sim-(T-T_c)^{-0.6975} +{\rm const.}
$]{\includegraphics[width=3.2in]{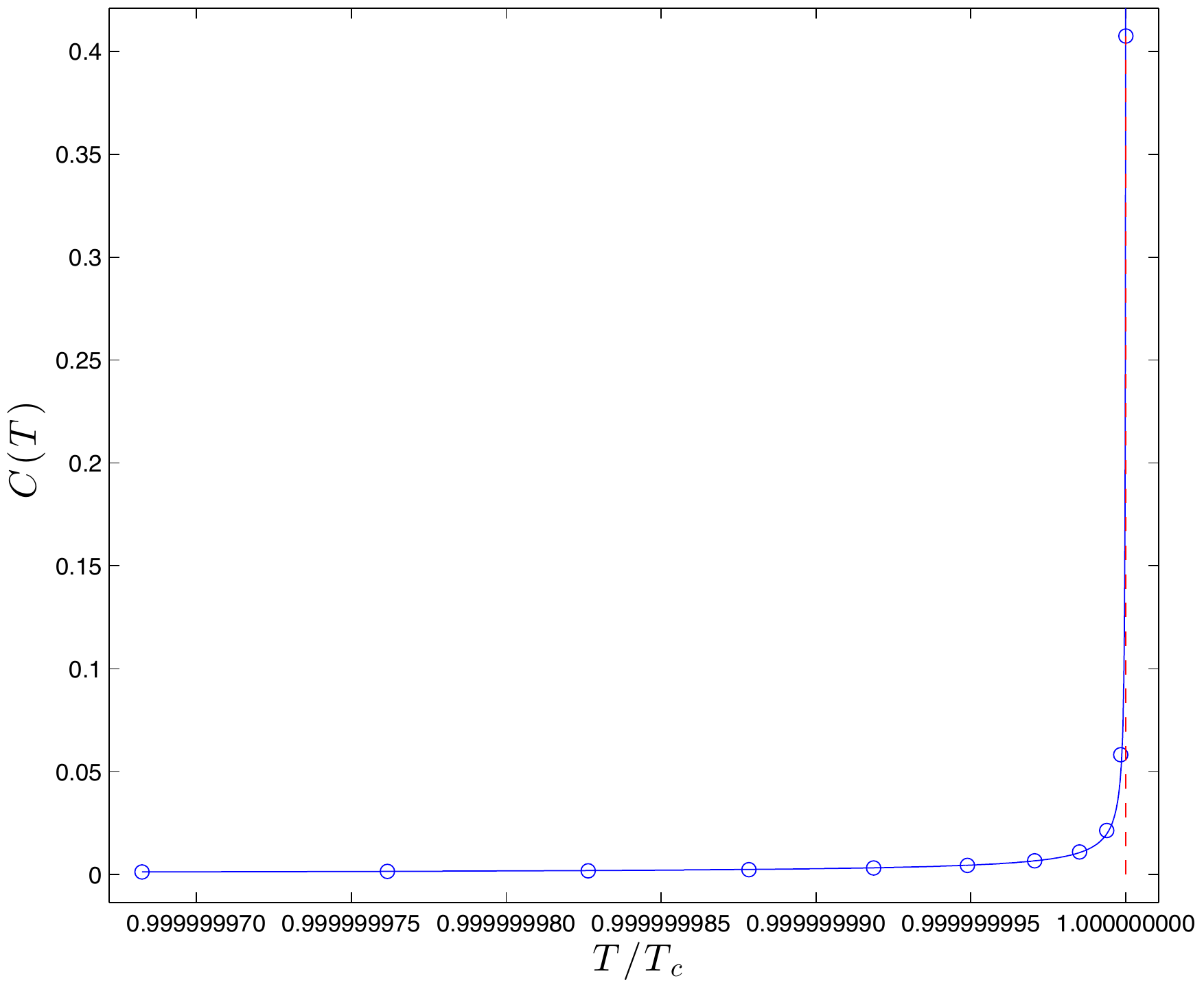}} 
\subfigure[Fit: $\log(C(T))=-0.6975\log(T_c-T) +{\rm const.}
$]{\includegraphics[width=3.1in]{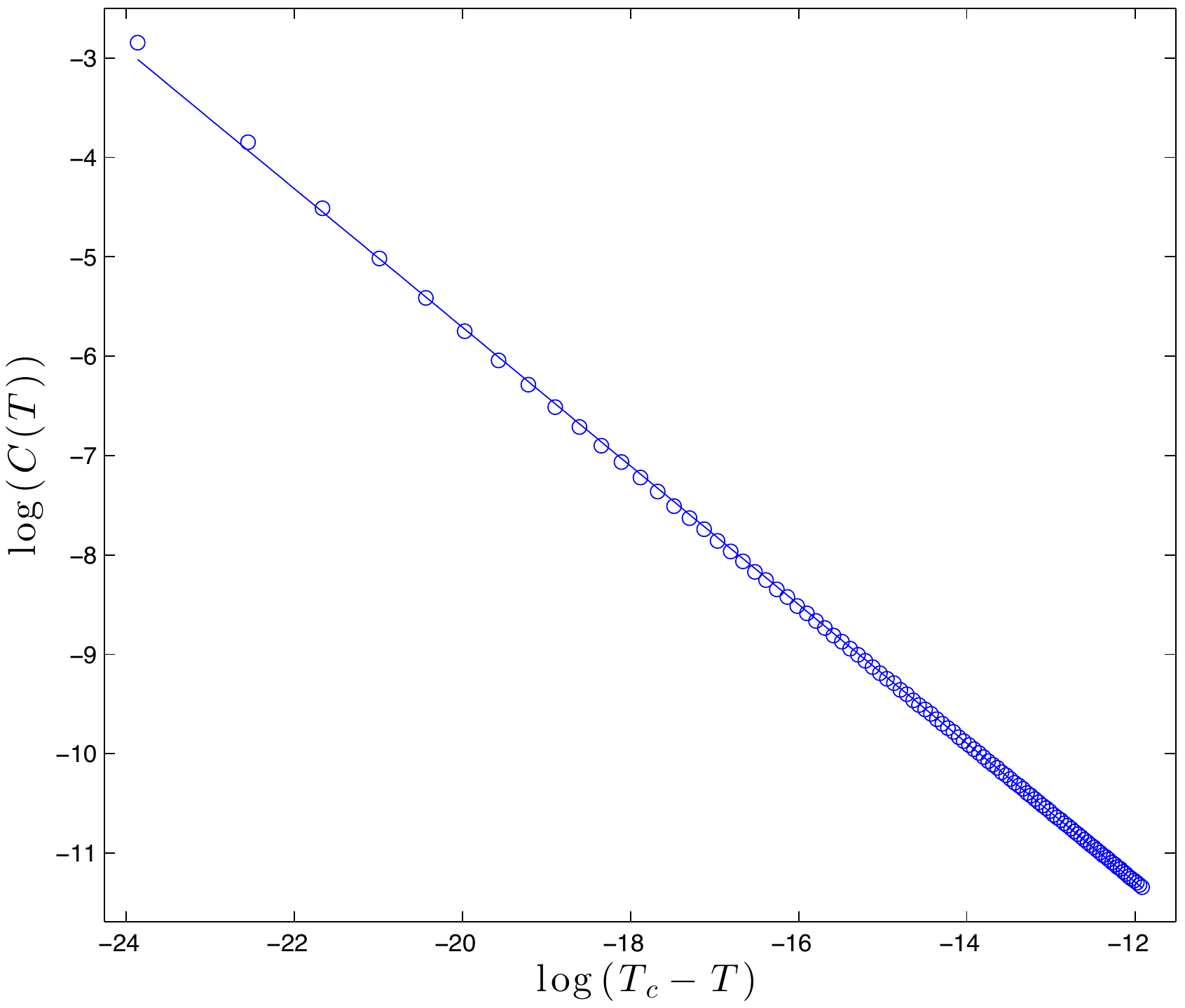}}
   \caption{\small The specific heat defined for the entanglement entropy {\it vs.} temperature along a path defined in the text. On the left is shown the fit for the ten nearest to $T=T_c$ of the computed points. The fit  to a critical exponent of $\sim 0.6975$  is shown to the right. See text for more discussion.}   \label{fig:specific_heat_crit}
   \end{center}
\end{figure}

Starting with the relation~(\ref{eq:betaess}) above for the black holes we can  examine the shape of the critical  ($q=q_c$) curve near the critical point at $\beta_c$ by writing $S_{BH}=S_{BH}^c+\epsilon$, where $S^c_{BH}=l^2\pi/6$,  and expanding in small $\epsilon$, giving the leading behaviour:
\begin{equation}
\label{eq:neighbourhood1}
\beta-\beta_c=-\frac{27}{2} \frac{\beta_c}{(\pi l^2)^3}(S_{BH}-S_{BH}^c)^3 +
 \hdots
\end{equation}
The coefficient and zero  are particular to the function black hole $S_{BH}(\beta)$, but what's important for us  is the cubic power, since it controls important universal behaviour\cite{Chamblin:1999hg}. It follows from it  that the black hole specific heat at constant charge $C_q^{\rm BH}(T)= T({\partial S_{\rm BH}}/{\partial T})_q$ has a critical exponent of 2/3: $C_q^{\rm BH}(T)\sim(T-T_c)^{-2/3}$. We might ask if the same sort of physics is present for the entanglement entropy $S(T)$, from which we can  define an analogous specific heat $C(T)=T(\partial S/\partial T)$. Indeed after subtracting off the temperature independent piece $S^\circ$ for convenience, the leading behaviour in the neighbourhood of the critical point is indeed fit by a cubic.  (There are of course, depending upon how closely one probes, subleading terms which could limit the numerical accuracy of the determination of the universal physics there.)   Critical behaviour can be explored directly numerically by computing  the entanglement entropy $S(T)$ at $q=q_c$ in the neighbourhood of $T_c$ at enough points to allow a reliable computation of the derivative, and hence $C(T)$.  We studied~$C(T)$ this way at  100 points to get an estimate of the critical exponent  (we chose an approach such that   $T/T_c-1\sim10^{-8}$) obtaining an approximate value of  0.6975.  See figure~\ref{fig:specific_heat_crit}. It is important to note that the number of significant digits listed here corresponds to the accuracy of the fit overall, but there's uncertainty in each point itself (each obtained from a complicated  boundary value numerical exercise). Taking this into account, we estimate that this means that to the accuracy we are working at, our result of $0.6975$ is roughly consistent with the exponent  $2/3$.  Proceeding to higher precision is very computationally intensive, but worth exploring further in case (for example)  there is some weak $\theta_0$ dependence.

Of course,  this particular critical point is perhaps not directly applicable to second order points of interest in (for example) condensed matter systems being currently considered in the literature, (note that further analysis of it in the fixed charge ensemble has shown\cite{Chamblin:1999hg} it to be in a basin of instability to fluctuations\footnote{Also, it is the fixed charge ensemble, so somewhat different character from other examples, although that alone does not prevent it from  finding a role somewhere. }), but it is  pleasing to be able to have such a tractable model of this kind of behavior of entanglement entropy in {\it any} system, and it may well be instructive behavior to have observed, and to study further. 

In that spirit, we can examine the behaviour of the entanglement entropy specific heat as we approach the critical point  $q=q_c$ along other paths $q(T)$ we might choose to move along in the $(q,T)$ plane.  (Looking at the phase diagram in figure~\ref{fig:phasediagram}, our 2/3 exponent came from moving to the critical point along a horizontal path.) For comparison, if this was for the black holes, we'd have by differentiating $S_{\rm BH}(T)$:
\begin{equation}
C^{\rm BH}(T)= \left(2S_{\rm BH}+\frac{\pi^{1/2}}{S_{\rm BH}^{1/2}}\frac{dq(T)}{dT}\right)
\left(\frac{S_{\rm BH}-\pi q^2+{3S_{\rm BH}^2}/{\pi l^2}}{3\pi q^2 +{3S_{\rm BH}^2}/{\pi l^2}-S_{\rm BH}}\right)\ ,
\end{equation}
(In a different thermodynamic ensemble, in which the role of the pressure  thermodynamic variable~$P$ is played by the cosmological constant (see {\it e.g.}, refs.\cite{Caldarelli:1999xj,Dolan:2011xt,Kubiznak:2012wp}),  this quantity reduces to $C^{\rm BH}_P(T)$.) Generically, the singularity of this function is controlled by the denominator of the right hand factor, which is the same polynomial whose double zero is located at the critical point (see equation~(\ref{eq:criticalcondition})). 

\begin{figure}[h]
\begin{center}
\subfigure[Fit: $C(T)\sim-(T-T_c)^{-9916} +{\rm const}$.]{\includegraphics[width=3.2in]{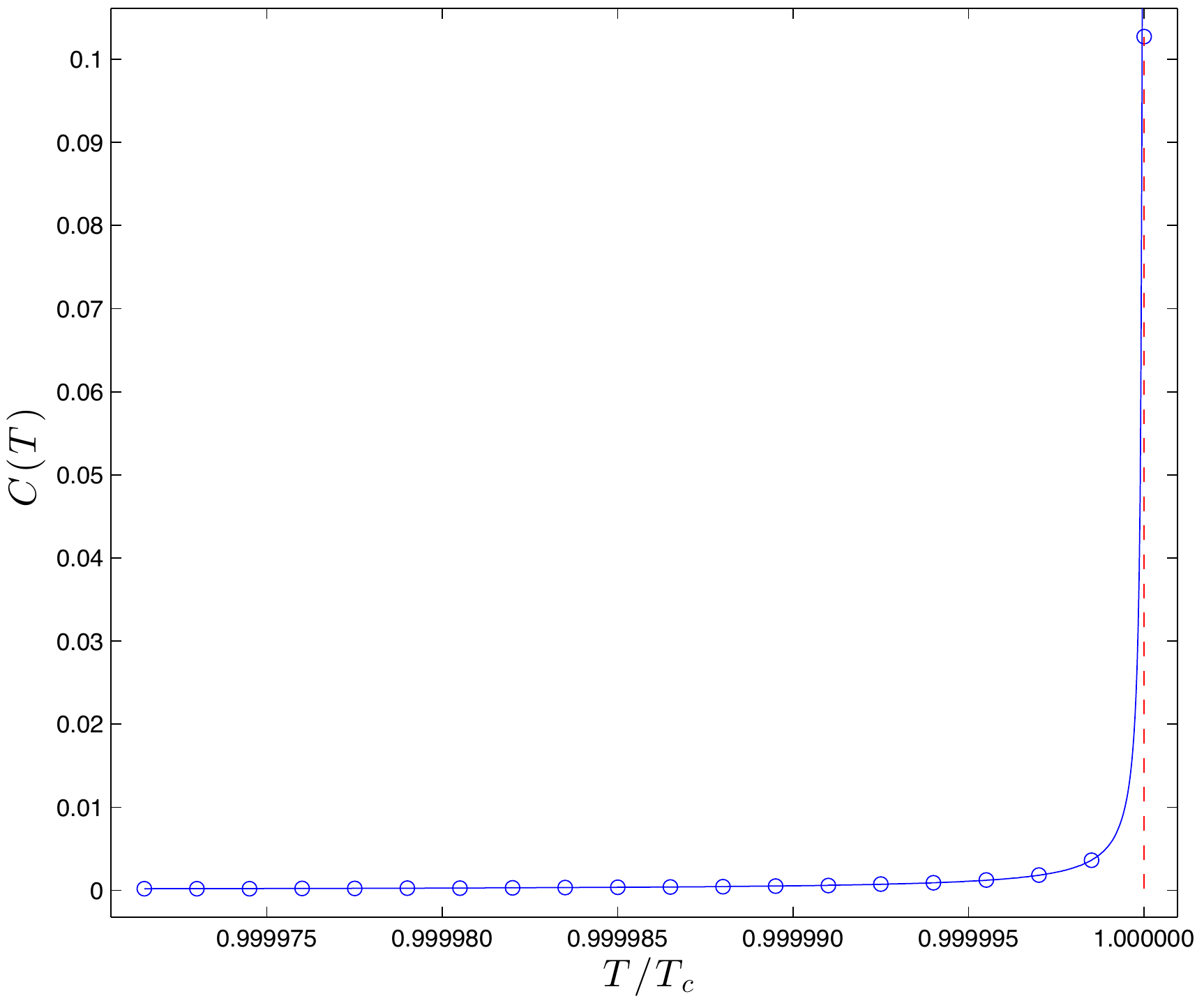}} 
\subfigure[Fit: $\log(C(T))=-0.9916\log(T_c-T) +{\rm const}$.]{\includegraphics[width=3.15in]{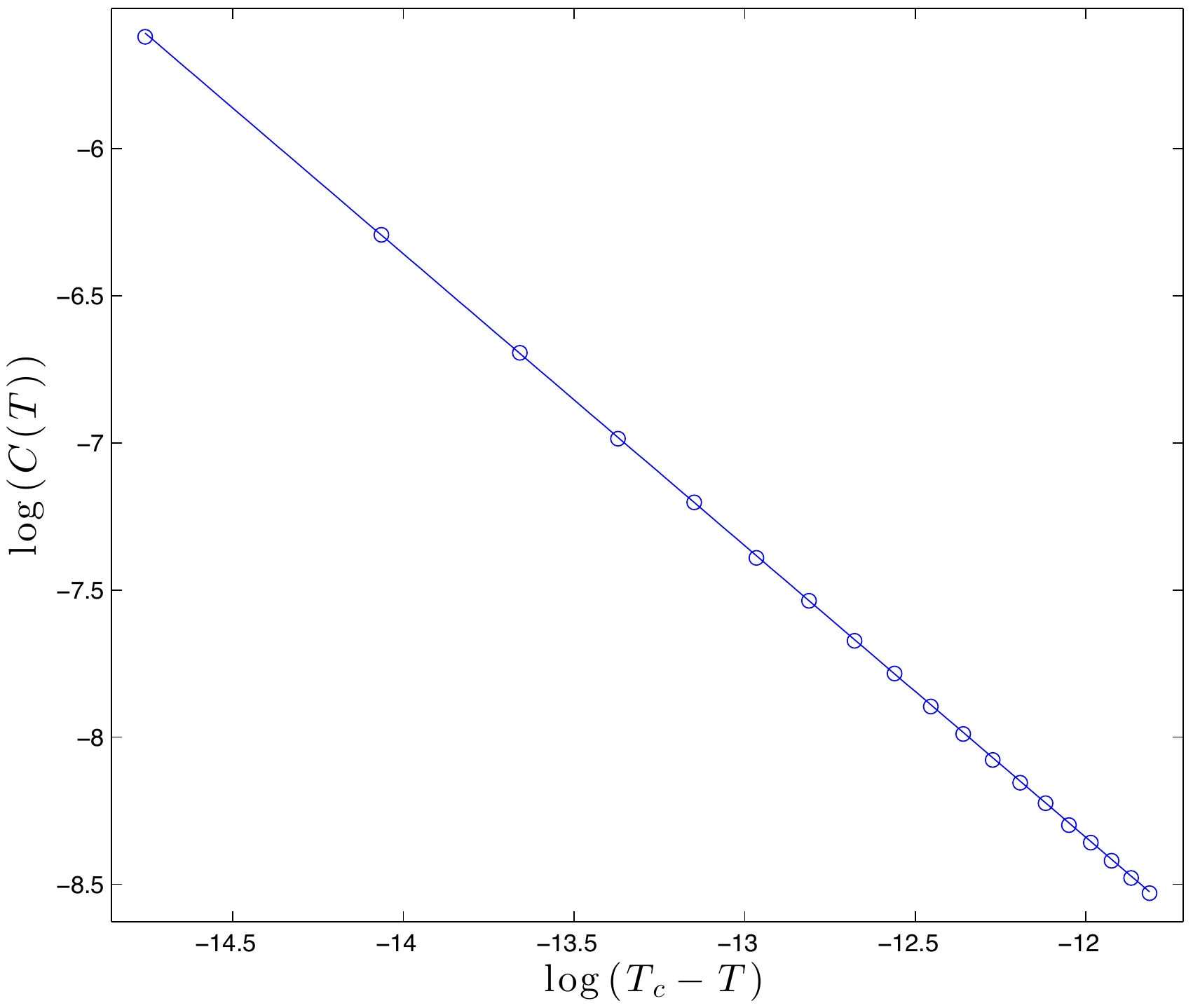}}
   \caption{\small The specific heat defined for the entanglement entropy {\it vs.} temperature along a path defined in the text, for 20 points. The fit consistent with a critical exponent of unity is shown to the right.}   \label{fig:specific_heat}
   \end{center}
\end{figure}

On  the approach to the critical point $C^{\rm BH}(T)$ diverges  with a critical exponent of unity: $C^{\rm BH}(T)\sim |T-T_c|^{-1}$.  This will be true for generically chosen paths $q(T)$ toward the critical point. (The special non--generic path $q=q_c$ is the one with exponent 2/3.) So we might wonder if this is true for our analogous specific heat $C(T)$ derived  from the entanglement entropy (again away from the regime where ${\cal A}$ is large and the quantity becomes the thermal entropy).  In another series of  computations  we explored a path to criticality by choosing a sequence  of $M$ charges very close to $q_c$ (we chose $q_m=q_c+ml/10^{6}$, where $q_c=l/6$ and $m=0,1,2,\ldots, M-1 $) and then at each $q_m$, working out the entanglement entropy $S(T)$ along the state curve for that charge for enough points to allow a reliable evaluation of a discrete version of $C(T)=T{\partial S}/{\partial T}$ to high accuracy. Running this close to criticality means that the derivative will be changing fast, and so a lot of computational effort needs to be expended. We chose around 2000 points for each $q_m$, and then $C(T)$ was evaluated at the point where the second derivative of  $\beta(r_+)$ vanishes (equivalently, choosing where $C$ is a maximum is also a good special point). This gives a well--defined sample path to criticality\footnote{In fact, if the phase diagram in figure~\ref{fig:phasediagram} were sufficiently magnified, the path would be seen arriving at the critical point  from above right, tangential to the line of first order points that extends  down and to the left.}, and we took  $M=20$ points along this path, ending on the critical point. The behaviour near $T_c$ (we had $T/T_c-1\sim 10^{-5}$) gave a good numerical fit  (a value of $0.9916$)  that was consistent with the power law $C \sim |T-T_c|^{-1}$.  See figure~\ref{fig:specific_heat}. (Similar cautionary remarks about the overall numerical accuracy, made earlier for the 2/3 critical exponent, apply here. Note, however, that to get this result we used fewer points along the path and were not as close to $T_c$, so this critical exponent was somewhat easier to establish than for  the fixed charge case.)

Such specific heats (defined with respect to temperature derivatives) control physics associated with thermal fluctuations, and it  would be interesting to go further and examine other, non--thermal quantities that can be derived from the entanglement entropy for these phase transitions. This is particularly important since the entanglement entropy may distinguish itself most as a  diagnostic tool  in the realm of quantum fluctuations. Pertinent to this has been some of the  discussion in the literature\cite{Bhattacharya:2012mi,Nozaki:2013vta} about the quantity that would play the role that temperature does in the first law as the conjugate of entropy. This could be an important quantity when considering systems at zero temperature, fixed temperature, or even away from thermodynamic equilibrium where there temperature may be ill--defined. The inverse size of region ${\cal A}$ is thought to  define that analogous temperature, and it would be interesting to study analogues of the specific heat defined using that quantity for the class of transitions we have studied here. We leave that for future work.

\section{Concluding Remarks}

By studying the holographic entanglement entropy in a finite volume system, a number of interesting phenomena became accessible. The finite nature of the system means that the entropy maximizes in an interesting way, rising to a certain value and then (by symmetry --- modulo horizon--wrapping contributions) falling again. Most interesting was the ready access one has to observing the behavior of the entanglement  entropy near important phase transitions that are present at finite volume (but thermodynamically  possible due to having a large number of degrees of freedom at large $N$). Recall that unless there is another physical scale against which to compare the temperature, such as a confinement scale like $\Lambda_{\rm QCD}$, or something analogous,  in infinite volume the transition temperature is driven to zero\footnote{See ref.\cite{Klebanov:2007ws} for an early study of holographic entanglement entropy through a deconfining phase transition in infinite volume.}, so finite volume allows for study of the neighbourhood of the transitions.
 The Hawking--Page first order transition\cite{Hawking:1982dh}  (and its finite potential generalization of ref.\cite{Chamblin:1999tk}) saw the expected jump in entanglement entropy due to the increase in the number of degrees of freedom, to a phase (controlled by large black holes) where the entropy rises most rapidly with temperature. The first order transitions between black holes in the fixed charge ensemble\cite{Chamblin:1999tk} were also seen to have  a jump in the entanglement  entropy, although it is necessarily smaller.  (It is of order pure numbers, as opposed to of order~$N^2$.)  The high temperature regime of the entanglement entropy for all cases  here is in some sense universal, being controlled by the large black hole behaviour, which is perhaps not too surprising.  

 The  setting allowed a study of the novel second order phase transition of ref.\cite{Chamblin:1999tk} as well, revealing a stationary point in the entanglement entropy itself, as a function of temperature --- a feature that (as far as we know) has not been seen before. This may be worth further study, to gain more insight into how entanglement entropy behaves across important second (and other) order phase transitions of relevance to various systems of interest theoretically and experimentally.  It was notable that the entanglement entropy's temperature dependence, even though we were far from the thermal regime  (${\cal A}$ was small), inherited  key features of the form of the temperature dependence of the parent black holes' entropy, including a critical point. We examined the physics in the neighbourhood of this point, and  to the accuracy that we worked,  we saw that the critical point has the same singular behaviour (critical exponents) for the specific heat as the black hole physics, a fact not guaranteed {\it \`a priori}. 
 
 It is worth noting that we can already contrast what we have seen here with results in ref.\cite{Albash:2012pd} for  the  entanglement entropy in a system with a second order transition, which was also  studied holographically.  That was for a fully back--reacted supergravity dual of a superconductor\cite{Bobev:2011rv}, and that was the first such study of the entanglement entropy of such a system. Interestingly, while again there was a change in the slope of the entropy as a function of $T$ across the transition, the change was {\it discontinuous}, while here it was smooth (through a point of inflection). Both situations are fully back--reacted gravity solutions, so it would be interesting to study further what features (beyond a condensing scalar and perhaps finite volume) account for the stark differences in behaviour.  This is probably an important class of examples to enlarge so that further study may be pursued.

\bigskip

\bigskip

{\bf Note:} As these results were being prepared for publication, a paper\cite{Hubeny:2013gta} appeared on the arxiv which may present results that coincide with some discussed here.

\section*{Acknowledgements}
 CVJ would like to thank the  US Department of Energy for support under grant DE-FG03-84ER-40168, Tameem Albash for useful conversations, Rob Myers and Mukund Rangamani each for  helpful conversations (after the first version of this manuscript appeared) and Amelia for her support and patience.


\begin{thebibliography}{10}

\bibitem{'tHooft:1973jz}
G.~'t~Hooft, ``A planar diagram theory for strong interactions,''
{\em Nucl. Phys.} {\bf B72} (1974)  461.

\bibitem{Maldacena:1997re}
J.~M. Maldacena, ``The large n limit of superconformal field theories and
  supergravity,'' {\em Adv. Theor. Math. Phys.} {\bf 2} (1998)  231--252,
\href{http://arxiv.org/abs/hep-th/9711200}{{\tt hep-th/9711200}}.

\bibitem{Gubser:1998bc}
S.~S. Gubser, I.~R. Klebanov, and A.~M. Polyakov, ``Gauge theory correlators
  from non-critical string theory,'' {\em Phys. Lett.} {\bf B428} (1998)
  105--114,
\href{http://arxiv.org/abs/hep-th/9802109}{{\tt hep-th/9802109}}.

\bibitem{Witten:1998qj}
E.~Witten, ``Anti-de sitter space and holography,'' {\em Adv. Theor. Math.
  Phys.} {\bf 2} (1998)  253--291,
\href{http://arxiv.org/abs/hep-th/9802150}{{\tt hep-th/9802150}}.

\bibitem{Witten:1998zw}
E.~Witten, ``Anti-de sitter space, thermal phase transition, and confinement in
  gauge theories,'' {\em Adv. Theor. Math. Phys.} {\bf 2} (1998)  505--532,
\href{http://arxiv.org/abs/hep-th/9803131}{{\tt hep-th/9803131}}.

\bibitem{Ryu:2006bv}
S.~Ryu and T.~Takayanagi, ``{Holographic derivation of entanglement entropy
  from AdS/CFT},'' \href{http://dx.doi.org/10.1103/PhysRevLett.96.181602}{{\em
  Phys.Rev.Lett.} {\bf 96} (2006)  181602},
\href{http://arxiv.org/abs/hep-th/0603001}{{\tt arXiv:hep-th/0603001
  [hep-th]}}.

\bibitem{Ryu:2006ef}
S.~Ryu and T.~Takayanagi, ``{Aspects of holographic entanglement entropy},''
  {\em JHEP} {\bf 08} (2006)  045,
\href{http://arxiv.org/abs/hep-th/0605073}{{\tt arXiv:hep-th/0605073}}.

\bibitem{Hawking:1982dh}
S.~W. Hawking and D.~N. Page, ``{Thermodynamics of Black Holes in anti-De
  Sitter Space},''
\href{http://dx.doi.org/10.1007/BF01208266}{{\em Commun. Math. Phys.} {\bf 87}
  (1983)  577}.

\bibitem{Chamblin:1999tk}
A.~Chamblin, R.~Emparan, C.~V. Johnson, and R.~C. Myers, ``Charged {A}d{S}
  black holes and catastrophic holography,'' {\em Phys. Rev.} {\bf D60} (1999)
  064018,
\href{http://arxiv.org/abs/hep-th/9902170}{{\tt hep-th/9902170}}.

\bibitem{Aharony:1999t}
O.~Aharony, S.~S. Gubser, J.~M. Maldacena, H.~Ooguri, and Y.~Oz, ``Large n
  field theories, string theory and gravity,'' {\em Phys. Rept.} {\bf 323}
  (2000)  183--386,
\href{http://arxiv.org/abs/hep-th/9905111}{{\tt hep-th/9905111}}.

\bibitem{Chamblin:1999hg}
A.~Chamblin, R.~Emparan, C.~V. Johnson, and R.~C. Myers, ``Holography,
  thermodynamics and fluctuations of charged ads black holes,'' {\em Phys.
  Rev.} {\bf D60} (1999)  104026,
\href{http://arxiv.org/abs/hep-th/9904197}{{\tt hep-th/9904197}}.

\bibitem{Kubiznak:2012wp}
D.~Kubiznak and R.~B. Mann, ``{P-V criticality of charged AdS black holes},''
  \href{http://dx.doi.org/10.1007/JHEP07(2012)033}{{\em JHEP} {\bf 1207} (2012)
   033},
\href{http://arxiv.org/abs/1205.0559}{{\tt arXiv:1205.0559 [hep-th]}}.

\bibitem{Gunasekaran:2012dq}
S.~Gunasekaran, R.~B. Mann, and D.~Kubiznak, ``{Extended phase space
  thermodynamics for charged and rotating black holes and Born-Infeld vacuum
  polarization},'' \href{http://dx.doi.org/10.1007/JHEP11(2012)110}{{\em JHEP}
  {\bf 1211} (2012)  110},
\href{http://arxiv.org/abs/1208.6251}{{\tt arXiv:1208.6251 [hep-th]}}.

\bibitem{Gubser:2008px}
S.~S. Gubser, ``{Breaking an Abelian gauge symmetry near a black hole
  horizon},'' \href{http://dx.doi.org/10.1103/PhysRevD.78.065034}{{\em
  Phys.Rev.} {\bf D78} (2008)  065034},
\href{http://arxiv.org/abs/0801.2977}{{\tt arXiv:0801.2977 [hep-th]}}.

\bibitem{Hartnoll:2008kx}
S.~A. Hartnoll, C.~P. Herzog, and G.~T. Horowitz, ``{Holographic
  Superconductors},''
  \href{http://dx.doi.org/10.1088/1126-6708/2008/12/015}{{\em JHEP} {\bf 12}
  (2008)  015},
\href{http://arxiv.org/abs/0810.1563}{{\tt arXiv:0810.1563 [hep-th]}}.

\bibitem{Herzog:2009xv}
C.~P. Herzog, ``{Lectures on Holographic Superfluidity and
  Superconductivity},''
  \href{http://dx.doi.org/10.1088/1751-8113/42/34/343001}{{\em J. Phys.} {\bf
  A42} (2009)  343001},
\href{http://arxiv.org/abs/0904.1975}{{\tt arXiv:0904.1975 [hep-th]}}.

\bibitem{Hartnoll:2009sz}
S.~A. Hartnoll, ``{Lectures on holographic methods for condensed matter
  physics},'' \href{http://dx.doi.org/10.1088/0264-9381/26/22/224002}{{\em
  Class.Quant.Grav.} {\bf 26} (2009)  224002},
\href{http://arxiv.org/abs/0903.3246}{{\tt arXiv:0903.3246 [hep-th]}}.

\bibitem{Faraggi:2007fu}
A.~Faraggi, L.~A. Pando~Zayas, and C.~A. Terrero-Escalante, ``{Holographic
  Entanglement Entropy and Phase Transitions at Finite Temperature},''
  \href{http://dx.doi.org/10.1142/S0217751X0904542X}{{\em Int. J. Mod. Phys.}
  {\bf A24} (2009)  2703--2728},
\href{http://arxiv.org/abs/0710.5483}{{\tt arXiv:0710.5483 [hep-th]}}.

\bibitem{Bah:2008cj}
I.~Bah, L.~A. Pando~Zayas, and C.~A. Terrero-Escalante, ``{Holographic
  Geometric Entropy at Finite Temperature from Black Holes in Global Anti de
  Sitter Spaces},'' \href{http://dx.doi.org/10.1142/S0217751X12500480}{{\em
  Int.J.Mod.Phys.} {\bf A27} (2012)  1250048},
\href{http://arxiv.org/abs/0809.2912}{{\tt arXiv:0809.2912 [hep-th]}}.

\bibitem{Fursaev:2006ih}
D.~V. Fursaev, ``{Proof of the holographic formula for entanglement entropy},''
  {\em JHEP} {\bf 09} (2006)  018,
\href{http://arxiv.org/abs/hep-th/0606184}{{\tt arXiv:hep-th/0606184}}.

\bibitem{Headrick:2007km}
M.~Headrick and T.~Takayanagi, ``{A holographic proof of the strong
  subadditivity of entanglement entropy},''
  \href{http://dx.doi.org/10.1103/PhysRevD.76.106013}{{\em Phys. Rev.} {\bf
  D76} (2007)  106013},
\href{http://arxiv.org/abs/0704.3719}{{\tt arXiv:0704.3719 [hep-th]}}.

\bibitem{Emparan:2006ni}
R.~Emparan, ``{Black hole entropy as entanglement entropy: A Holographic
  derivation},'' \href{http://dx.doi.org/10.1088/1126-6708/2006/06/012}{{\em
  JHEP} {\bf 0606} (2006)  012},
\href{http://arxiv.org/abs/hep-th/0603081}{{\tt arXiv:hep-th/0603081
  [hep-th]}}.

\bibitem{Blanco:2013joa}
D.~D. Blanco, H.~Casini, L.-Y. Hung, and R.~C. Myers, ``{Relative Entropy and
  Holography},''
\href{http://arxiv.org/abs/1305.3182}{{\tt arXiv:1305.3182 [hep-th]}}.

\bibitem{Casini:2011kv}
H.~Casini, M.~Huerta, and R.~C. Myers, ``{Towards a derivation of holographic
  entanglement entropy},''
  \href{http://dx.doi.org/10.1007/JHEP05(2011)036}{{\em JHEP} {\bf 1105} (2011)
   036},
\href{http://arxiv.org/abs/1102.0440}{{\tt arXiv:1102.0440 [hep-th]}}.

\bibitem{Hubeny:2007xt}
V.~E. Hubeny, M.~Rangamani, and T.~Takayanagi, ``{A covariant holographic
  entanglement entropy proposal},''
  \href{http://dx.doi.org/10.1088/1126-6708/2007/07/062}{{\em JHEP} {\bf 07}
  (2007)  062},
\href{http://arxiv.org/abs/0705.0016}{{\tt arXiv:0705.0016 [hep-th]}}.

\bibitem{Hubeny:2012wa}
V.~E. Hubeny and M.~Rangamani, ``{Causal Holographic Information},''
  \href{http://dx.doi.org/10.1007/JHEP06(2012)114}{{\em JHEP} {\bf 1206} (2012)
   114},
\href{http://arxiv.org/abs/1204.1698}{{\tt arXiv:1204.1698 [hep-th]}}.

\bibitem{Bekenstein:1973ur}
J.~D. Bekenstein, ``{Black holes and entropy},''
\href{http://dx.doi.org/10.1103/PhysRevD.7.2333}{{\em Phys.Rev.} {\bf D7}
  (1973)  2333--2346}.

\bibitem{Bekenstein:1974ax}
J.~D. Bekenstein, ``{Generalized second law of thermodynamics in black hole
  physics},''
\href{http://dx.doi.org/10.1103/PhysRevD.9.3292}{{\em Phys.Rev.} {\bf D9}
  (1974)  3292--3300}.

\bibitem{Hawking:1974sw}
S.~Hawking, ``{Particle Creation by Black Holes},''
\href{http://dx.doi.org/10.1007/BF02345020}{{\em Commun.Math.Phys.} {\bf 43}
  (1975)  199--220}.

\bibitem{Hawking:1976de}
S.~Hawking, ``{Black Holes and Thermodynamics},''
\href{http://dx.doi.org/10.1103/PhysRevD.13.191}{{\em Phys.Rev.} {\bf D13}
  (1976)  191--197}.

\bibitem{Caldarelli:1999xj}
M.~M. Caldarelli, G.~Cognola, and D.~Klemm, ``{Thermodynamics of
  Kerr-Newman-AdS black holes and conformal field theories},''
  \href{http://dx.doi.org/10.1088/0264-9381/17/2/310}{{\em Class.Quant.Grav.}
  {\bf 17} (2000)  399--420},
\href{http://arxiv.org/abs/hep-th/9908022}{{\tt arXiv:hep-th/9908022
  [hep-th]}}.

\bibitem{Dolan:2011xt}
B.~P. Dolan, ``{Pressure and volume in the first law of black hole
  thermodynamics},''
  \href{http://dx.doi.org/10.1088/0264-9381/28/23/235017}{{\em
  Class.Quant.Grav.} {\bf 28} (2011)  235017},
\href{http://arxiv.org/abs/1106.6260}{{\tt arXiv:1106.6260 [gr-qc]}}.

\bibitem{Bhattacharya:2012mi}
J.~Bhattacharya, M.~Nozaki, T.~Takayanagi, and T.~Ugajin, ``{Thermodynamical
  Property of Entanglement Entropy for Excited States},''
  \href{http://dx.doi.org/10.1103/PhysRevLett.110.091602}{{\em Phys.Rev.Lett.}
  {\bf 110} (2013)  091602},
\href{http://arxiv.org/abs/1212.1164}{{\tt arXiv:1212.1164}}.

\bibitem{Nozaki:2013vta}
M.~Nozaki, T.~Numasawa, A.~Prudenziati, and T.~Takayanagi, ``{Dynamics of
  Entanglement Entropy from Einstein Equation},''
\href{http://arxiv.org/abs/1304.7100}{{\tt arXiv:1304.7100 [hep-th]}}.

\bibitem{Klebanov:2007ws}
I.~R. Klebanov, D.~Kutasov, and A.~Murugan, ``{Entanglement as a probe of
  confinement},'' \href{http://dx.doi.org/10.1016/j.nuclphysb.2007.12.017}{{\em
  Nucl.Phys.} {\bf B796} (2008)  274--293},
\href{http://arxiv.org/abs/0709.2140}{{\tt arXiv:0709.2140 [hep-th]}}.

\bibitem{Albash:2012pd}
T.~Albash and C.~V. Johnson, ``{Holographic Studies of Entanglement Entropy in
  Superconductors},'' \href{http://dx.doi.org/10.1007/JHEP05(2012)079}{{\em
  JHEP} {\bf 1205} (2012)  079},
\href{http://arxiv.org/abs/1202.2605}{{\tt arXiv:1202.2605 [hep-th]}}.

\bibitem{Bobev:2011rv}
N.~Bobev, A.~Kundu, K.~Pilch, and N.~P. Warner, ``{Minimal Holographic
  Superconductors from Maximal Supergravity},''
  \href{http://dx.doi.org/10.1007/JHEP03(2012)064}{{\em JHEP} {\bf 1203} (2012)
   064},
\href{http://arxiv.org/abs/1110.3454}{{\tt arXiv:1110.3454 [hep-th]}}.

\bibitem{Hubeny:2013gta}
V.~E. Hubeny, H.~Maxfield, M.~Rangamani, and E.~Tonni, ``{Holographic
  entanglement plateaux},''
\href{http://arxiv.org/abs/1306.4004}{{\tt arXiv:1306.4004 [hep-th]}}.

\end{thebibliography}

\providecommand{\href}[2]{#2}\begingroup\raggedright\endgroup

\end{document}